\documentclass[a4paper,11pt]{article}
\usepackage[utf8]{inputenc}
\usepackage[T2A]{fontenc}
\usepackage{authblk}
\usepackage{blindtext}
\usepackage[russian, english]{babel}
\usepackage{amsmath,amssymb}
\usepackage{cite}
\usepackage[unicode,linktocpage=true,plainpages=false,pdfpagelabels=false]{hyperref}

\newcommand{\dd}{\partial}
\newcommand{\de}{\delta}
\newcommand{\al}{\alpha}
\newcommand{\be}{\beta}
\newcommand{\ka}{\varkappa}
\newcommand{\la}{\lambda}
\newcommand{\pp}{\varphi}
\newcommand{\ep}{\varepsilon}
\newcommand{\ga}{\gamma}
\newcommand{\Ga}{\Gamma}
\newcommand{\simup}[2]{\left\{#1\right\}^{#2}}
\newcommand{\simdown}[2]{\left\{#1\right\}_{#2}}
\newcommand{\antisimup}[2]{\left[#1\right]^{#2}}

\newcommand{\gMim}{\bar{g}}

\newcommand{\D}{\nabla}

\allowdisplaybreaks

\begin{document}
\author[ ]{R.~V.~Ilin \thanks{E-mail:~\href{mailto:rireilin@gmail.com}{rireilin@gmail.com}}}
\title{Noether currents in theories with higher derivatives and theories with differential field transformation in action (DFTA)}
\affil[$\ast$]{{\it Saint Petersburg State University, Saint-Petersburg, Russia}}
\date{}
\maketitle

    \begin{abstract}
        The article is devoted to the investigation of the Noether currents and integrals of motion in the special subclass of theories with higher field derivatives - theories under differential field transformations in action (DFTA). Under fairly general assertions, we derive a simple representation for the integrals as sums of "old" integrals of motion, terms, that vanishes on the "old" equations of motion and surface integrals. We show that for some cosmological theories of that kind with high order derivatives the last contribution can violate gauge invariance, specifically, diffeomorphisms. Then we investigate ambiguity of the Noether procedure for higher derivative theories and fix it in the way the problem resolves in quite general setting. The obtained results are discussed for the several extensions of the General Relativity, which are obtained by DFTA, in particular, for mimetic, disformal and Regge-Teitelboim theories of gravity. In particular, we calculate Iyer-Wald black hole entropy from the corrected currents and also discuss the derivation of the integral constraints on Sachs-Wolfe effect.
    \end{abstract}

    \vskip 9em

\section{Introduction}
For the last two decades, much effort has been made to solve the problems of dark energy and dark matter. Despite the numerous approaches \cite{NojiriOdintsovModifiedGravReview2011} (see also recent report \cite{ModifiedGravNutshellNojiri2017}) that were proposed to solve these problems, still none of them can be treated by now as the final one. Such stagnation over the years has inevitably caused the complexity of all newly proposed models to grow. Compared to the past years, now it is not uncommon to consider theories with high order derivatives for the mentioned purposes (see, for example, review \cite{IshakTestingGenRelCosm2018}). However, one is most likely to have many issues with the analysis of the theory of this kind due to the complex nature of the resulting equations of motion. Probably the most notorious one is linear instability that can be easily seen from the Hamiltonian formalism (see \cite{ChenOstrogradkyGhost2013} and references therein). That said, it is even more useful, if some of the properties of the usual theories with no more than first field derivatives in the action are preserved for theories with higher orders. One such property is the Noether's theorem \cite{noether}. Conserved quantities, especially energy-momentum, can provide shortcuts for some physical properties of the theory, sometimes bypassing the above-mentioned difficulties. For instance, it is well-known that there is a strong connection between Noetherian charges and black hole entropy (see details, for example, in the seminal work \cite{IyerWaldCovariantPS}). One might also expect the connection between Noether's energy-momentum and Hamiltonian similarly to the case of the usual theories with no more than first derivatives.\par

In this article we investigate several important issues that can accompany Noether currents in the general case of theories with high order field derivatives. Here we are focused on the original definition, that have been proposed in the famous work of Noether \cite{noether} but sill widely used (see, for example, \cite{PetrovCovariantized, EmtsovaPetrovToporenskyTele1, EmtsovaPetrovToporenskyTele2, TothNoetherTeukolsky2018, ObukhovPotralesPuetzfeldConsCurrents2015} and also \cite{CapozzielloLaurentisNoetherCosmology2012}). It is also much simpler than some others (see, for example, \cite{BarnichGeneralForm}) and is better suited for generalizations. Even for this definition the general case is too complex. However, such scale of analysis is somewhat unnecessary for the problems that the average high-order theory has. For that reason we will work mainly with the specific class of them - with theories, that are obtained by differential transformation in action (DFTA).\par

Though these theories are not uncommon (tetrad gravity, Palatini approach, theories obtained by Weyl transform and etc.), they were recognized as useful cosmological tools only recently. The key feature of these theories is the existence of the additional degrees of freedom while those of the "initial", non-modified theory are untouched \cite{statja60} and present in theory with change. The most popular models of this kind are mimetic gravity \cite{mukhanov2014} and its various modifications \cite{MomeniNewModifiedMimetic,mimetic-review17,HorndeskiMimetic,NorijiOdintsovGhostFreeMimetic2017,OdintsovNojiriMimetic2014, OdintsovOikonomouMimeticAcceleratingCosmologies2016, MimeticMONDLikeVagnozzi, MimeticAliveAndWell2018, CovarHoravaMimeticHorndesky2016, F2AbelianMimeticGorjiMukhoyama, F2YangMillsMimeticGorjiMukhoyama}. Another example is provided by Regge-Teitelboim embedding gravity \cite{regge, tapia, pavsic85, deser, rojas04} (it is also sometimes called geodetic brane theory, see, for example \cite{davkar}). For the purposes of the current article theories with DFTA are mainly interesting because nearly all of them are high-order by construction: if one starts with theory that have at least first order derivatives of the fields $\la$, then after the change $\la \equiv \la(\pp, \dd\pp)$, the new action will in general depend on the second derivatives of $\pp$. As it will be shown below, the change on its own greatly simplifies the form of the Noether's current and its superpotential (in gauge theories), which allows one to establish some important properties of the current in a very general setup. One such property that is specific to the case of DFTA is reducibility of Noether's current of the resulting theory to the current of the "old", unchanged theory, if it is evaluated on the "old" solutions. In that sense, theories with DFTA can be used as litmus test of the conserved current's construction procedure. For the solutions of the original theory, not all combinations of new fields are observable, and therefore meaningful integrals of motion should reflect this property. If they still depend on these quantities, it will be an indicator that the definition of conserved quantities is incorrect for the theory (and probably the whole class of them) and needs to be modified.\par

One can also study conserved currents for theories with DFTA to clarify some physics behind the extra degrees of freedom. It was shown \cite{pavsic85, statja60} that for the mentioned theories one can write equations of motion as a system that consists of Einstein's equations with certain energy-momentum tensor $\tau_{\mu\nu}$ and also of equations that constraints it. In the simple case of mimetic gravity, it can be seen \cite{mukhanov} that this tensor corresponds to the pressureless dust with 4-velocity $u_\mu \equiv \dd_\mu\sigma$ where scalar $\sigma$ is one of the new independent fields introduced by DFTA. Another field in this theory - auxiliary metric $\gMim_{\mu\nu}$ - is not observable since it is shadowed by the physical metric $g_{\mu\nu}$ in the equations of motion. These facts allow one \cite{Golovnev201439} to ditch auxiliary metric and reformulate theory as GR with the scalar matter restricted by normalization condition $g^{\mu\nu}\dd_\mu\sigma\dd_\nu\sigma = 1$. In more complex cases such as Regge-Teitelboim gravity, the exact relation between $\tau_{\mu\nu}$ and new fields can be extremely difficult to establish. In these situations, integrals of motion can hint at some properties of the matter in cases for which equations constraining $\tau_{\mu\nu}$ are too complex to analyze. Conserved quantities obtained by Noether's procedure are also of the special interest due to their close relationships with canonical approach.\par

It is well-known that for the several gauge theories - namely, generally covariant - the currents may have the additional problems. The most notorious among them is non-covariance of the conserved currents and consequent unability to correctly define corresponding charges \cite{misner, landavshic2, Baryshev, julia, BabakGrischukEnergy}. This problem has a long history and has been studied by many different authors in the context of General Relativity and other gauge theories of gravity (see, for example, discussions in \cite{Baryshev, julia} and references therein). Usually the main idea comes down to the fact that for many generally covariant theories Noether's theorem does not give the suitable energy-momentum tensor - because, in fact, it is not a tensor at all. One possible solution is to use the current that depends on the gauge parameters $\xi^a$ and then evaluate conserved quantities by choosing the appropriate ones \cite{Komar}. The choice of $\xi^a$ is usually based on the boundary conditions (see, for example, \cite{BarnichGeneralForm, AveryNoetherWard, julia, katz, linden-bell}). However, for GR with the action that depends on the first metric derivatives the corresponding current (it was obtained in \cite{katz, ChrushcielSuperpotential1988, julia}) is also non-covariant in sense that it is not a vector density, which is even worse than non-tensorial behavior of energy-momentum tensor. Though, at the moment there are several approaches to construction of the specific superpotentials, which lead to physically meaningful results \cite{BarnichGeneralForm, BakCangemiJakiw, julia, juliaPart2, linden-bell} under certain assumptions, the non-covariance of Katz superpotential does not seem so unnatural. After all, it corresponds to the action with no more than first derivatives, which is known to be non-invariant under diffeomophisms on boundary. Unfortunately, it is not completely clear if the current is guaranteed to be vector density or not in the general case of high-order derivative theory.
Some work has been done in this directions for theories with second derivatives allowed in the action (see, for example, \cite{PetrovCovariantized, PetrovLompayTekinKopeikin}), but in general the answer is still unknown due to the complexity of the analysis required. In this article we will show that for the above-mentioned mimetic gravity the diffeomophisms current is not a vector density even when the action is diffeomorphisms-invariant.\par

The outline of the paper is the following. In the section \ref{GeneralCascadeSection} we provide brief overview of the Noether's theorem for theories with high-order field derivatives with (gauge) symmetry. At this point we do not restrict neither the number of derivatives of the gauge parameters in the transformation laws, nor the form of the Lagrangian. By following the procedure similar to the one proposed in \cite{WaldGeneralSuperpotential}, (see also \cite{TothNoether'sTheorem2017}), the current of the gauge theory is rewritten as the divergence of the superpotential (2-form). In the section \ref{MimeticSuperpotentialSubsection} we apply these results to the specific high-order theories with DFTA - mimetic gravity and Regge-Teitelboim theory. We show that the results obtained are physically unacceptable for various reasons: for mimetic gravity corresponding current is not vector density and also does not reduce to the Komar superpotential \cite{Komar} on the GR solutions, while for the Regge-Teitelboim gravity the current is zero for any solution.\par

In the section \ref{TheoriesWithChangeSection} we move from the most general case to the theories with DFTA. The primary interest here is how currents and superpotentials transform under DFTA. Unlike previous section, here the cases under consideration have restrictions on derivative orders of both the {\it original} fields in action $N_o$ and the maximal order of derivatives of gauge parameters in the fields' transformation law $M$. These restrictions, however, do not preclude any modern physically interesting theory of such kind. The subsection \ref{FirstOrderChangeSubsection} is devoted to the most general gauge-invariant theory with $N_o = 1, M = 1$ and DFTA with the maximal order of derivatives of new fields in it $W = 1$. For this case, we derive simple expression for the current as a sum of the three terms: conserved current of the theory without transformation, terms that vanish on the its equations of motion, and a certain divergence of the superpotential. We show that such current transforms as a vector density in cases when fields in the original theory lie in the tensor representation of the symmetry group. The subsection \ref{ScalarTensorChangeSubsection} generalizes these results to the more interesting case $N_o = 2, M = 1$.\par
As discussed in \cite{PetrovCovariantized}, the construction of Noether's integrals of motion in high-order derivative theories is ambiguous even when using the standard formulae. Roughly speaking, it is due to the possibility of the ordering derivatives in the expression for the current. It is worth noting that this ambiguity is much more restrictive than the above-mentioned possibility to add an exact form to the conserved current. In the subsection \ref{AmbiguitySection} we discuss this problem in detail for the theories with DFTA. We show, that for the cases considered in \ref{FirstOrderChangeSubsection} and \ref{ScalarTensorChangeSubsection} there exists a special way to fix this ambiguity. The resulting current is identical to that of the original theory modulo the original equations of motions. If the original current was vector density, and certain conditions are imposed on the transformation formula, then the new current appears to be vector density too. In the section \ref{Applications} these general results are applied to specific models: mimetic and disformal theories\cite{DisformalBekenstein, DisformalMimDeruelleRua} gravities, Regge-Teitelboim theory and also their Lagrange multiplier reformulations. Physical quantities such as total mass, angular momentum are discussed in this context. Moreover, we apply corrected conservation laws to the well-known derivation of Iyer-Wald entropy and argue that if the change of variables contain scalar fields, there will be no new contributions to the BH entropy compared to the original theory. For example, in mimetic gravity the Iyer-Wald entropy for black holes can be computed by using the same Komar 2-form. Finally, the end of this section is devoted to discussion of how corrected superpotentials can be used to derive integral constraints on Sachs-Wolfe effect in theories with change.\par
Finally, in the section \ref{Conclusion} we briefly conclude and propose directions for the further investigation.

\section{Notations and abbreviations}\label{Notations}
We will always call independent field arguments of the action "independent variables" of the theory similar to the arguments of functions. In that sense one may also say that the independent variables of the theory are the variables one must vary independently according to the least action principle to obtain all the equations of motion. Another term that will be used for them is simply "fields". We remind that we will use abbreviation "DFTA" for differential field transformations. In some cases we will also use the word "change" for it.\par
For the sake of convenience when working with the total symmetrization or antisymmetrization over arbitrary many indices, we use slightly different notations compared to the standard ones. For some quantity $A^{\mu\nu}$ we have the following notations:
\begin{align}
    &\antisimup{A^{\mu\nu}}{\mu\nu} \equiv A^{\mu\nu} - A^{\nu\mu},\label{AntisymmetrizerDef}\\
    &\simup{A^{\mu\nu}}{\mu\nu} \equiv A^{\mu\nu} + A^{\nu\mu},
    \label{SymmetrizerDef}
\end{align}
Here the indices related to the figure (square) brackets are {\it not} coordinate or gauge indices. They simply denote over which indices one should (anti-)symmetrize the expression within the brackets.
For quantities with more than two indices, the definition is analogous: after the figure (square) bracket one writes all indices, over which the expression within it should be fully (anti-)symmetrized. Note that the normalization factors in these definitions are absent.\par
In some situations we will use the following extended notation for the product of the partial derivative operators with free indices:
\begin{equation}
    \dd_{\al_m\dots\al_k}\equiv
    \begin{cases}
        \dd_{\al_m}\dd_{\al_{m+1}}\dots\dd_{\al_k},\;\; m\leq k,\\
        I,\;\; m = k + 1,\\
        0,\;\; m > k + 1,
    \end{cases},
    \label{dirDefinition}
\end{equation}
where $I$ is the identity operator. We will also use similar $dots$ notation to {\it extend} indexed objects:
\begin{equation}
    \tilde{S}^{\al_m\dots\al_k}\equiv
    \begin{cases}
        S^{\al_m\dots\al_k},\;\; m\leq k,\\
        0,\;\; m > k + 1.
    \end{cases}
    \label{TensorDotsIndicesNotations}
\end{equation}
These notation are not meant to replace the standard ones - they are only useful for few scenarios. Their main benefit throughout the article is to allow one to include "boundary" terms in sums without any need to write them separately. For example, for some coordinate function $A(x)$ one has:
\begin{equation}
    \sum_{i = 0}^k\tilde{S}^{\al_2\dots\al_i}\dd_{\al_2\dots\al_i}A = SA+S^{\al_2}\dd_{\al_2}A+\dots+S^{\al_2\dots\al_k}\dd_{\al_2}\dots\dd_{\al_k}A.
\end{equation}
Note that here we omitted the term $i = 0$ because both $S^{\al_2\dots \al_0}$ and $\dd_{\al_2\dots\al_0}$ are zero. For simplicity we will further {\it omit tilde} and suppose that all objects with indices with subscript are {\it extended} by means of \eqref{TensorDotsIndicesNotations}.\par
We will assume, that the greek indices take values $0,..., D-1$. Small Latin indices are used to denote the gauge indices unless otherwise specified. For brevity, we use big Latin indices to denote multiindices of the form $A = \mu_1..\mu_k a_1..a_s$.\par
When working with local identities it is common to use the jet space $J^\infty\left(M,\mathbb{R}\right)$ derivatives of some functions (in our cases -- of the lagrangian). In its simplest form, this jet space can be represented as the product $M\times V^\infty$ where $V^k$ is a Taylor polynomials space with the coordinates $\pp_A,\dd_\mu\pp_A,...\dd_{\mu_1..\mu_k}\pp_A$, and $M$ is space-time. We will use the following notation for the Lagrangian derivatives:
\begin{equation}
    Z_{\left(\pp\right)}^{A\mu_1\dots\mu_k} \equiv \frac{\dd L}{\dd\dd_{\mu_1\dots\mu_k}\pp_A}.
    \label{ZNotion}
\end{equation}
Reader not familiar with the jet bundles may think of these objects as of usual derivatives of $L$ with respect to the arguments $\dd_{\mu_1..\mu_k}\pp_A$, while the other derivatives of fields including just fields must be considered independent arguments of $L$. This definition is enough for the understanding the paper, as we are interested primarily in the local aspects of the conserved currents.\par
To denote equations that are satisfied on-shell, instead of the standard sign $"{}="$ we will use the sign $"\approx"$ as it is accepted in the related literature.\par
In the discussion of the integrals of motion it is convenient to use differential form notation for the currents $J^\rho$. Though the currents usually are vectors for theories in the flat space-time, the situation can be more complex when it has curvature. For these cases the current is usually expected to be vector density (see, for example, \cite{Komar, PetrovCovariantized}), however, as we can see futher, it can be violated by the presence of high-order derivatives of fields in the action. Following the standard conventions (see, for example, \cite{BarnichGeneralForm}), we will treat by definition $(\sqrt{-g})^{-1}J^\rho$ as the components of $1$-form $\star J$ if $J^\rho$ is a vector density.

\section{Cascade equations: brief review}\label{GeneralCascadeSection}
In the current section, we briefly discuss the application of Noether's theorems to the gauge field theories with high-order field derivatives in the action. Nearly all of the results of these sections are not ideologically new in contrast to the subsequent ones. For example, one may find alternative derivation of the expression for the superpotential in \cite{WaldGeneralSuperpotential}, also see discussion in \cite{TothNoether'sTheorem2017}. However, the explicit formulae are not commonly mentioned in the literature. They are very useful for further discussion, so we spent some time to re-derive them without going into much detail.\par
Our main interest here revolves around theories on a space-time D-dimensional manifold $M$ with independent fields $\pp_A$ and the Lagrangian $L$, which locally depends on $\pp_A$ and its derivatives up to the order $N$:
\begin{equation}
    L = L(\pp_A,\dots,\dd_{\al_1\dots\al_N}\pp_A).
\end{equation}
The action of such theory is constructed in straightforward manner:
\begin{equation}
    S = \int_M d^Dx L.
\end{equation}
One can use space-time metric among the fields $\pp_A$. It may also be included as background field, which slightly changes the analysis of conserved currents as it is shown in the section \ref{TheoriesWithChangeSection}.\par
We will assume, that the action of such theory has symmetry under some gauge group in the sense, that its variation with respect to the infinitesimal gauge transformation is a surface term:
\begin{equation}
    \de S = -\int_{M} d^Dx\dd_\mu K^\mu\left(\xi^a,\dots,\dd_{\al_1\dots\al_C}\xi^a\right),
    \label{ActionVariationGeneralCase}
\end{equation}
where $\xi^a$ denotes the gauge parameter. Though the currents for our main examples will correspond to diffeomophisms, all the general cases in this section and the subsequent ones are analyzed for the arbitrary gauge theory. In that regard, we will use the following parametrization for the variation of the fields under the mentioned gauge transformation:
\begin{equation}
    \de\pp_A \equiv -\sum_{i = 0}^{M}H_{(i)Aa}{}^{\al_1\dots\al_i}\dd_{\al_1\dots\al_i}\xi^a.
    \label{fieldVarGeneralFormula}
\end{equation}
Though in the definition \eqref{ActionVariationGeneralCase} r.h.s may seem unnecessary, it allows one to consider many viable field theories, such as General Relativity with action that depends on no more than first metric derivatives, in the same manner as those with strictly gauge-invariant actions.\par
It should be noted, that until the end of the article we will use the standard derivatives instead of the covariant ones. The reason behind this is the generality of the approach. Firstly, covariant derivatives in the Lagrangian are very useful in cases when the fields $\pp_A$ behave like tensors under the gauge transformations. In the work \cite{WaldGeneralSuperpotential} it was shown, that this fact leads to many useful properties for the conserved current and the corresponding superpotential. However, if the transformation law of the $\pp_A$ is more complex, one should pick all relevant gauge-invariant combinations as the arguments of the Lagrangian. This approach can be helpful in particular situations but not in the general case. Secondly, in some cases, it is suitable to violate gauge invariance only on the boundary, while preserving it in the volume. The original formulation of TEGR (see, for example, \cite{GolovnevTeleIntro}) or general relativity with first-order derivative action and independent metric provide good examples of this situation. In these situations the Lagrangian is manifestly non-gauge-invariant, so the the choice of its arguments in the tensor form is impossible.\par

 \subsection{Local identities}\label{LocalIdentitiesSubsection}
 To derive Noether's currents we will go the standard way described, for example, in the works \cite{linden-bell, PetrovLompayTekinKopeikin, julia, juliaPart2} (see also Appendix in \cite{BakCangemiJakiw}).
 The formula \eqref{ActionVariationGeneralCase} defines the corresponding variation of the Lagrangian under the infinitesimal gauge transformations with parameters $\xi^a$. On the other hand, this variation is defined by the variations of its arguments, which together with \eqref{ActionVariationGeneralCase} leads to the following:
 \begin{equation}
     -\dd_\mu K^\mu = \sum_{i = 0}^{N}Z^{A\al_1\dots\al_i}\dd_{\al_1\dots\al_i}\de\pp_A.
     \label{NoetherTheoremStep1}
 \end{equation}
 Recall that hereinafter we use the notation \eqref{ZNotion}. Expressing in this formula the term with $i = 0$ by using the definition of the variational derivative:
 \begin{equation}
     \sum_{i = 0}^{N}(-1)^j\dd_{\al_1\dots\al_i}Z^{A\al_1\dots\al_i} = \frac{\de S}{\de \pp_A},
 \end{equation}
 and also Leibniz rule:
\begin{multline}
    Y^{\al_1\dots\al_j}\dd_{\al_1\dots\al_j}Q
    +(-1)^{j+1}(\dd_{\al_1\dots\al_j}Y^{\al_1\dots\al_j})Q=\\
    =\dd_{\rho}\Bigg[\sum_{i=0}^{j-1}(-1)^{i+j+1}(\dd_{\al_1\dots\al_{j-i-1}}Y^{\rho\al_1\dots\al_{j-1}})(\dd_{\al_{j-i}\dots\al_{j-1}} Q)\Bigg],
    \label{Leibniz'Rule}
\end{multline}
where $Y^{\al_1\dots\al_j}$ is arbitrary and fully-symmetric, one can derive from \eqref{NoetherTheoremStep1} the relation, that is a "local" conservation law if the equations of motion are applied:
\begin{equation}
    \dd_{\rho}J^{\rho}=-\frac{\de S}{\de \pp_A}\de \pp_A,
    \label{OffShellTotalNoetherIdentity}
\end{equation}
where
\begin{equation}
    J^{\rho} \equiv \sum_{j = 1}^N\sum_{i = 0}^{j-1}(-1)^{i+j+1}\dd_{\al_1\dots\al_{j-i-1}}Z^{A\rho\al_1\dots\al_{j-1}}\dd_{\al_{j-i}\dots\al_{j-1}}\de\pp_A+K^\rho.
    \label{JDefinition}
\end{equation}
Algebraic Poincare lemma (see, for example, the section from \cite{BarnichBrandtHanneauxBRST} and the references within) states, that any {\it identity} of the form:
\begin{equation}
    \dd_\mu A^\mu \equiv 0,
    \label{IdenticalConservationLaw}
\end{equation} 
where $A^\mu$ is a local vector density, automatically implies that $A^\mu$ is a divergence of the local superpotential $V^{\be\rho}$:
\begin{equation}
    A^\mu \equiv \dd_\nu V^{\nu\mu},\;\;\;\; V^{\be\rho} = -V^{\rho\be},
    \label{AlgebraicPoincareLemma}
\end{equation}
If the r.h.s of the \eqref{OffShellTotalNoetherIdentity} was zero, we could use this theorem and consider the superpotential $V^{\al\be}$ instead of the current $J^\rho$. In the general case, it cannot be done. Nevertheless, locality of the symmetry allows one to consider quantity $J'^\rho$ related to $J^\rho$ by a simple formula, which will be "identically conserved" in the sense of \eqref{IdenticalConservationLaw}. Then one can use lemma \eqref{AlgebraicPoincareLemma} and present $J^\rho$ in the following form:
\begin{equation}
    J^\rho = S^\rho+\dd_\be V^{\be\rho},
    \label{SVCurrentDecomposition}
\end{equation}
where $S^\rho$ vanishes on the equations of motion. The latter of the subsection to the large extent is devoted to the derivation of this representation. In some cases it will be analogous to the proof from \cite{WaldGeneralSuperpotential} (see also \cite{Samokhvalov2021}). The only difference here is the fact, that in \cite{WaldGeneralSuperpotential} the fields' transformation laws are restricted, while here we do not impose any additional conditions on them. It should be also noted, that for the sake of simplicity algebraic Poincare lemma will not be used here; instead we will rely on the stronger condition, namely on the fact, that the relation \eqref{OffShellTotalNoetherIdentity} holds for the arbitrary choice of $\xi^\mu$.\par
As the first step towards the proof of decomposition \eqref{SVCurrentDecomposition} we should bring \eqref{OffShellTotalNoetherIdentity} to the form of the "identical" conservation law. Consider variation of the action under gauge transformation with the infinitesimal gauge parameter $\xi^a$ which has a compact support:
\begin{equation}
    \de S = \int_M d^Dx\left(\sum_{i = 0}^M\frac{\de S}{\de\pp_A}H_{(i)Aa}{}^{\al_1\dots\al_i}\dd_{\al_1\dots\al_i}\xi^a\right).
\end{equation}
Conditions imposed on $\xi^a$ allows one to integrate by parts for free and then use arbitrariness of $\xi^a$ and also that one of its support:
\begin{equation}
    \sum_{i = 0}^{M}(-1)^i\dd_{\al_1\dots\al_i}\left(\frac{\de S}{\de \pp_A}H_{(i)A\mu}{}^{\al_1\dots\al_i}\right) = 0.
    \label{SecondNoether'sTheorem}
\end{equation}
These identities are well-known in gauge theories as the statement of the "Second Noether's theorem" (see, for example, \cite{noether,BarnichGeneralForm, AveryNoetherWard}) or just "Bianchi identities" as mentioned in the introduction. Let us express the first term with $i = 0$ through the rest of the sum and then substitute it in the r.h.s. of the \eqref{OffShellTotalNoetherIdentity}:
\begin{equation}
    \dd_{\rho}J^{\rho} = \sum^{M}_{i = 1}\left(\frac{\de S}{\de \pp_A}H_{(i)A\mu}{}^{\al_1\dots\al_i}\dd_{\al_1\dots\al_i}\xi^{\mu}+(-1)^{i+1}\dd_{\al_1\dots\al_i}\left(\frac{\de S}{\de \pp_A}H_{(i)A\mu}{}^{\al_1\dots\al_i}\right)\xi^{\mu}\right).
\end{equation}
Here we again use Leibniz' rule \eqref{Leibniz'Rule} and finally get "strong" conservation law:{}
\begin{equation}
    \dd_{\rho}J'^{\rho} = 0,
    \label{J'ConsLaw}
\end{equation}
where
\begin{align}
    &J'^{\rho} \equiv J^{\rho} + X^{\rho},
    \label{J'Definition}\\
    &X^{\rho} \equiv \sum_{j = 1}^M\sum_{i = 0}^{j-1}(-1)^{i+j}\dd_{\al_1\dots\al_{j-i-1}}\left(\frac{\de S}{\de \pp_A}H_{(j)Aa}{}^{\rho\al_1\dots\al_{j-1}}\right)\dd_{\al_{j-i}\dots\al_{j-1}}\xi^a.
    \label{XDefinition}
\end{align}
The obtained formula \eqref{J'ConsLaw} is linear combination of $\xi^a$ and its derivatives with maximum derivative order of $N+M$. Thus we may use the following decomposition for the $J'$:
\begin{equation}
    J'^{\rho} = -\sum_{k = 0}^{N_{d}}K_{(k)a}{}^{\rho\al_1..\al_k}\dd_{\al_1\dots\al_k}\xi^a,
    \label{J'KDecomposition}
\end{equation}
where $N_{d} \equiv N+M-1$, and the quantities $K_{(k)a}{}^{\rho\al_1..\al_k}$ are assumed to be fully symmetric with respect to the indices $\al_1..\al_k$. One can choose $\xi^a$ and its derivatives independently at one point, so this relation can be further decomposed into the system of identities:
\begin{align}
    &\dd_{\rho}K_{(0)}{}_{a}{}^{\rho} = 0,\label{KDecompositionReccurentChain0}\\
    &\dd_{\rho}K_{(k)}{}_{a}{}^{\rho\al_1\dots\al_k}+\frac{1}{k!}\left\{K_{(k-1)}{}_{\al}{}^{\al_1..\al_k}\right\}^{\al_1\dots\al_k}=0,\;\;\;\; 1\leq k \leq N_{d}
    \label{KDecompositionReccurentChain1}\\ &\left\{K_{(N_{d})a}{}^{\al_1\dots\al_{N_{d}+1}}\right\}^{\al_1\dots\al_{N_{d}+1}} = 0.
\label{KDecompositionReccurentChain2}
\end{align}
The analysis of this chain of equations will allow one to represent $J^\rho$ in the form \eqref{SVCurrentDecomposition}. To start, note that the symmetry of $K_{(k)}{}_{a}{}^{\rho\al_1..\al_k}$ over the indices $\al_1..\al_k$ implies, that the following identity is satisfied:
\begin{equation}
    K_{(k-1)a}{}^{\al_k\al_1..\al_{k-1}} = \frac{1}{k!}\simup{K_{(k-1)a}{}^{\al_1\dots\al_{k}}}{\al_1\dots\al_k}+\frac{1}{k}\sum_{i = 1}^{k-1}\antisimup{K_{(k-1)a}{}^{\al_{k}\al_1\dots\al_{k-1}}}{\al_k\al_i},
    \label{SymmetricAntisymmetricPatrsDecomposition}
\end{equation}
where we used the notations \eqref{AntisymmetrizerDef}, \eqref{SymmetrizerDef}. The system of equations \eqref{KDecompositionReccurentChain0}-\eqref{KDecompositionReccurentChain2} technically is not a recurrent chain. However, similar methods can be used to find a new useful representation for $K_{(k)}{}_{a}{}^{\rho\al_1\dots\al_k}$. To do it we use \eqref{SymmetricAntisymmetricPatrsDecomposition} in the relation \eqref{KDecompositionReccurentChain1} and then act on the both sides of the resulting equation by the operator $(-1)^{k}\dd_{\al_{p+1}\dots\al_k}$. By summing the result with respect to $k$ from $p$ to $N_d$ one should get the following:
\begin{multline}
    \sum_{k=p}^{N_d}(-1)^{k}\dd_{\al_{p+1}\dots\al_{k+1}}K_{(k)}{}_{a}{}^{\al_{k+1}\al_1\dots\al_k} + \sum_{k=p}^{N_d}(-1)^{k}\dd_{\al_{p+1}\dots\al_k}K_{(k-1)a}{}^{\al_k\al_1\dots\al_{k-1}} =\\= \sum_{k=p+1}^{N_d}(-1)^{k}\frac{1}{k}\sum_{i = 1}^{k-1}\antisimup{\dd_{\al_{p+1}\dots\al_k}K_{(k-1)a}{}^{\al_k\al_1\dots\al_{k-1}}}{\al_k\al_i}.
    \label{DiagonalSumm}
\end{multline}
The l.h.s of this equations is obviously the telescopic series. By reassigning the indices $p\rightarrow p+1$ one may finally move to the appropriate expression for $K_{(p)a}{}^{\al_{p+1}\al_1..\al_{p}}$:
\begin{align}
    K_{(p)a}{}^{\al_{p+1}\al_1..\al_{p}}=\sum_{k=p}^{N_d}\sum_{i = 1}^{k}\frac{(-1)^{k-p}}{k+1}\antisimup{\dd_{\al_{p+2}..\al_{k+1}}K_{(k)a}{}^{\al_{k+1}\al_1..\al_{k}}}{\al_{k+1}\al_i},
    \label{KRepresentation}
\end{align}
This formula generalizes the analogous representation for $K_{(p)a}{}^{\al_{p+1}\al_1..\al_{p}}$ from \cite{ilinPastonUniverse}. In this work it was used to prove, that $K_{(1)\mu}{}^{\rho\be}$ is actually a superpotential (non-antisymmetric) for the energy-momentum tensor (in the current notation it just coincides with the $K_{(0)\mu}{}^{\rho}$). In the full analogy with the discussion of the case $N = 2, M = 1$ made in \cite{PetrovLompayTekinKopeikin}, we can substitute \eqref{KRepresentation} into \eqref{J'Definition} in order to find the corresponding superpotential for $J'$. After long calculations, that are presented in the appendix \ref{DerivationOfJ'FormulaAppendix}, the resulting expression for $J'^\rho$ takes the form:
\begin{equation}
J'^{\rho} = -\dd_{\be}\Bigg[\sum_{p = 1}^{N_d}\frac{p}{p+1}\antisimup{K_{(p)a}{}^{\rho\be\al_1..\al_{p-1}}}{\rho\be}\dd_{\al_1..\al_{p-1}}\xi^a+S^{\rho\be}\Bigg],
\label{J'AsDIvergenceWithNon-AntisymmetricSuperpotential}
\end{equation}
where
\begin{align}
\nonumber
    S^{\rho\be} =& \sum_{p = 2}^{N_d}\sum_{k= 1}^{p-1}\frac{k(-1)^{p-k}}{p+1}\dd_{\al_k..\al_{p-1}}\left(\antisimup{K_{(p)a}{}^{\rho\be\al_1..\al_{p-1}}}{\rho\al_{p-1}}+\antisimup{K_{(p)a}{}^{\be\rho\al_1..\al_{p-1}}}{\be\al_{p-1}}\right)\dd_{\al_1..\al_{k-1}}\xi^a-\\
\nonumber
    &-\sum_{p=3}^{N_d}\sum_{k = 1}^{p-1}(-1)^{p-k}\frac{k(k-1)}{p+1}\dd_{\al_{k}..\al_{p-1}}\antisimup{K_{(p)a}{}^{\al_{p-1}\rho\be\al_1..\al_{p-2}}}{\al_{p-1}\al_1}\dd_{\al_1..\al_{k-1}}\xi^a-\\
    &-\sum_{p=2}^{N_d}\frac{p(p-1)}{p+1}\antisimup{K_{(p)a}{}^{\be\rho\al_1..\al_{p-1}}}{\be\al_{p-1}}\dd_{\al_1..\al_{p-1}}\xi^a.
    \label{NonSymmetricAddition}
\end{align}
Despite the fact, that $J'{}^\rho$ now is a full divergence, the object under this divergence in the formula \eqref{J'AsDIvergenceWithNon-AntisymmetricSuperpotential} is not anti-symmetric. Hopefully, this fact originates only from $S^{\rho\be}$, so one can use the ambiguity of the expression under full divergence to fix this:
\begin{equation}
    \dd_\be S^{\rho\be} = \dd_\be \left(S^{\rho\be}+\dd_\al W^{\al\be\rho}\right),\;\;\;\; W^{\al\be\rho} = -W^{\be\al\rho}.
    \label{SuperpotSWAmbiguity}
\end{equation}
One may use the following expression for $W^{\al\be\rho}$ in order to anti-symmetryze $S^{\rho\be}$:
\begin{align}
    \dd_{\al}W^{\al\be\rho} \equiv -\sum_{p = 2}^{N_d}\dd_{\al_{p-1}}\Bigg[\sum_{k = 1}^{p-1}(-1)^{p-k}\frac{k(k+1)}{p+1}\dd_{\al_k..\al_{p-2}}\antisimup{K_{(p)a}{}^{\be\rho\al_1..\al_{p-1}}}{\be\al_{p-1}}\dd_{\al_1..\al_{k-1}}\xi^a\Bigg],
\end{align}
so, the expression under the divergence in r.h.s of \eqref{SuperpotSWAmbiguity} yields:
\begin{equation}
    S^{\rho\be}+\dd_\al W^{\al\be\rho} = \sum_{p = 2}^{N_d}\sum_{k = 1}^{p-1}\frac{k(-1)^{p-k}}{p+1}\dd_{\al_k..\al_{p-1}}\antisimup{K_{(p)a}{}^{\rho\be\al_1..\al_{p-1}}}{\rho\be}\dd_{\al_1..\al_{k-1}}\xi^a.
\end{equation}
By substituting this formula instead of $S^{\rho\be}$ in \eqref{J'AsDIvergenceWithNon-AntisymmetricSuperpotential} and by merging the resulting sums into one, we can write the final expressions for $S^\rho$ and $V^{\be\rho}$ from the decomposition \eqref{SVCurrentDecomposition}:
\begin{align}
    &S^{\rho}\equiv - X^{\rho}-\dd_{\be}\Bigg[\sum_{p = 1}^{N_d}\sum_{k = 1}^p\frac{k(-1)^{p-k}}{p+1}\dd_{\al_k..\al_{p-1}}W_{(p)a}{}^{\rho\be\al_1..\al_{p-1}}\dd_{\al_1..\al_{k-1}}\xi^a\Bigg]^{\rho\be},
    \label{SDef}\\
    &V^{\be\rho} = \Bigg[\sum_{p = 1}^{N_d}\sum_{k = 1}^{p}\frac{k(-1)^{p-k}}{p+1}\dd_{\al_k..\al_{p-1}}L_{(p)a}{}^{\be\rho\al_1..\al_{p-1}}\dd_{\al_1..\al_{k-1}}\xi^a\Bigg]^{\rho\be},
    \label{AntisymmetricSuperpotential}
\end{align}
where quantities $L_{(p)a}{}^{\rho\be\al_1..\al_{p-1}}$ and $W_{(p)a}{}^{\rho\be\al_1..\al_{p-1}}$ are the coefficients in the expansions for $J^\rho$ and $X^\rho$:
\begin{align}
    J^{\rho} = -\sum_{k = 0}^{N_{d}}L_{(k)a}{}^{\rho\al_1..\al_k}\dd_{\al_1..\al_k}\xi^{a}, \label{JLDecomposition}\\
    X^{\rho} = -\sum_{k = 0}^{N_{d}}W_{(k)a}{}^{\rho\al_1..\al_k}\dd_{\al_1..\al_k}\xi^{a}.\label{XDecomposition}
\end{align} 
It is worth noting that unlike the results of \cite{WaldGeneralSuperpotential}, there are no guarantees that the current $J^\rho$ corresponding to \eqref{SDef},\eqref{AntisymmetricSuperpotential} will be gauge invariant even if the lagrangian is a scalar density. One example of such a situation is provided by already mentioned mimetic gravity, which is briefly discussed below.

\subsection{Conserved charges}\label{ChargesSubsection}
If the current $J^\rho$ and its superpotential $V^{\rho\be}$ are vector and tensor densities respectively, one can use differential form notations to rewrite the relation \eqref{SVCurrentDecomposition}:
\begin{equation}
    J = S+dV, 
    \label{JFormDecomposition}
\end{equation}
where forms $S$ and $V$ can be easily calculated from $V^{\be\rho}$ and $S^\rho$, which are given by \eqref{SDef} and \eqref{AntisymmetricSuperpotential}. Recall (see section \ref{Notations}) that in the current notation $J, S$ and $V$ are $(D-1)-, (D-1)-$ and $(D-2)-$forms respectively. Relation \eqref{JFormDecomposition} allows one to use the standard procedures to define conserved charges from the currents $J^\rho$. The discussion below is more or less trivial generalization of the argument about the boundary conditions proposed in \cite{julia}. In general, the idea of constructing correct integrals of motion based on "background" quantities on the spatial infinity is very old, see for example \cite{katz, linden-bell, hawking, cornishEnergy, BarnichGeneralForm}.\par
For simplicity, we assume that the space-time in the theory under consideration is suitable for the decomposition on the timelike surfaces of the constant time $\Sigma_t$. Consider a surface, that contains a time-like direction of the form $B_r\equiv[t_2,t_1]\times s_r$, where $s_r$ denotes $(D-2)$-dimensional sphere of the radius $r$. We can integrate $J$ over this surface on-shell and then use the Stokes' theorem:
\begin{equation}
    \int_{B_\infty}J \approx \int_{s_\infty}V\Big|_{\Sigma_2}-\int_{s_\infty}V\Big|_{\Sigma_1}.
    \label{JFormStokesTheorem}
\end{equation}
If the integral in l.h.s was equal to zero, we would define the conserved charges as the following:
\begin{equation}
    Q[\pp_A,\xi_a] = \int_{S_\infty}V.
    \label{ChargeDef}
\end{equation}
To fulfill the stated condition and eliminate the l.h.s in the relation \eqref{JFormStokesTheorem} one can always impose {\it ad hoc} restrictions on the fields in addition to the least action principle. However, these conditions arise more naturally from the correctness of the least action principle on the boundary and also from the conditions on $K^\mu$ (see definition of the current \eqref{JDefinition}). Indeed, in the general form, the action variation can be written as such:
\begin{equation}
    \de S = \int_M d^Dx \frac{\de S}{\de\pp_A}\de\pp_A+\int_{\dd M} d^{D-1}\hat{x}\sqrt{\ga}n_\mu\alpha^\mu[\pp_A,\de\pp_A,...],
    \label{GeneralActionVar}
\end{equation}
where $\ga_{ij}$ is induced metric on the $\dd M$, $n_\mu$ denotes the normal to $\dd M$, and $\hat{x}$ are the local coordinates on $\dd M$. In order for the variational principle to be correct, the second term in the r.h.s of this relation has to vanish. This integral consists of two contributions. The first one comes from the timelike hypersurfaces $\Sigma_2$ and $\Sigma_1$, and the second one is the contribution from $B_\infty$. By definition of the least action principle, the fields and their time derivatives up to the order $N-1$ are fixed at the initial and final timelike hypersurfaces. Thus, the correctness condition comes down to the following:
\begin{equation}
    \int_{B_\infty}d^{D-1}\hat{x}\sqrt{\ga}n_\mu\alpha^\mu[\pp_A,\de\pp_A,...] = 0.
    \label{ActionPrincipleConsistency}
\end{equation}
Given the discussion at the beginning of the previous subsection, it is obvious that the first term in \eqref{JDefinition} and $\al^\mu$ are equal to each other. Finally, we impose the asymptotic condition on $K^\mu$ at space-like infinity:
\begin{equation}
    K^\mu = O\left(r^{-2-\ep}\right),\;\;\; \ep>0,
    \label{KAsymptoticCondition}
\end{equation}
where $A^\mu$ does not depend on r. Thus, the formula \eqref{ChargeDef} indeed defines the conserved charges. It is important to note, that the conditions \eqref{ActionPrincipleConsistency} restrict the set of functions $\pp_A$ that are valid for the variational principle. In turn, it also restricts the gauge parameter $\xi^b$ or, more precisely, its asymptotic behaviour at the spatial infinity. \par

\section{Noether superpotentials for mimetic gravity and Regge-Teitelboim theory}\label{MimeticSuperpotentialSubsection}
As mentioned earlier, the formula \eqref{AntisymmetricSuperpotential} does not guarantee the currents $J^\rho$ to be vector densities on the gauge group or to be just tensors from its representation. This problem was encountered before, see, for example, the discussion in \cite{katz}, \cite{julia} and also historical review \cite{Baryshev}. However, in such cases the most common explanation is the fact, that the Lagrangian of the theory isn't a scalar density (or just scalar in case of inner symmetries). So, the non-covariance of the currents in these cases is not surprising at all. As it was announced in the introduction, the analysis in the next sections is mostly devoted to theories with DFTA. One of the reasons to consider them here alongside with simplicity of the analysis is that for some theories from this class the current is not a vector density while the corresponding action is gauge-invariant.\par
One of the most popular theories with DFTA of this kind is mimetic gravity, firstly proposed in \cite{mukhanov}. Since then, it has received huge development in the works \cite{mukhanov2014, Golovnev201439, MomeniNewModifiedMimetic, mimetic-review17, ChamseddineAsymptoticallyFreeMimetic}. In its original formulation, mimetic gravity has a standard Einstein-Hilbert action with the simple scalar-tensor change:
\begin{equation}
    g_{\mu\nu} = \gMim_{\mu\nu}\gMim^{\al\be}\dd_\al\sigma\dd_\be\sigma,
    \label{MimeticChange}
\end{equation}
where $\gMim_{\mu\nu}$ is auxiliary metric, and $\sigma$ is scalar. To construct the conserved charges for this model firstly the superpotential \eqref{AntisymmetricSuperpotential} should be calculated. For mimetic gravity the first step leads to the expression for the superpotential:
\begin{multline}
    V_{Mim}^{\be\rho} = V_K^{\be\rho} + \frac{(D-1)}{3\ka}\sqrt{-g}\Bigg[\D^\rho\left(\D_\mu\sigma\D^\be\sigma\xi^{\mu}\right)+\\
    +\frac{1}{2}g^{\be\ep}\Ga^\rho_{\ep\la}\D_\mu\sigma\D^\la\sigma\xi^\mu+\frac{1}{2}\Ga^\rho_{\la\tau}g^{\la\tau}\D^\be\sigma\D_\mu\sigma\xi^{\mu}\Bigg]^{\rho\be},
    \label{MimeticSuperpotential}
\end{multline}
where $V_K^{\be\rho}$ is half of the well-known Komar superpotential \cite{Komar}. As can be seen from the expression, the extra terms that arose from DFTA violate the tensor density transformation law. This fact makes conserved charges dependent on the choice of coordinates, which is in fact a big problem. In some works (see, for example, \cite{julia, linden-bell, katz, ChrushcielSuperpotential1988}) it is proposed to use the background parameters such as auxiliary background metric to add certain terms to the current and therefore fix its transformation law. These parameters can either be purely external to the theory or appear from fixing the boundary conditions like those considered in the subsection \ref{ChargesSubsection}. There is also a possibility to "covariantize" the currents by using the ambiguity in Noether's procedure in theories with high-order derivatives, which was discussed in \cite{PetrovCovariantized}.\par
Apart from the tensor density transformation law, one may also require expression \eqref{MimeticSuperpotential} to reduce to Komar superpotential (which results from \eqref{AntisymmetricSuperpotential} for Einstein-Hilbert action) on the GR solutions. This behavior is expected, because, as it is mentioned above, the field transformation in action preserves all the solutions of the original theory. However, one may easily check, that \eqref{MimeticSuperpotential} does not satisfy this condition. Let physical metric describe spatially flat FLRW-universe. In this case one can use the following values for $\gMim_{\al\be}$ and $\sigma$:
\begin{equation}
    \gMim_{00} = 1,\;\; \gMim_{0i} = 0,\;\; \gMim_{ik} = -a^2(t)\ga_{ik},\;\; \sigma = x^0,
\end{equation}
where $\ga_{ik}$ denotes 3-dimensional euclidean metric, $a(t)$ is cosmological scale factor. Substituting this parametrization into \eqref{MimeticSuperpotential} leads to the following result for $V_{Mim}^{i0}$:
\begin{equation}
    V_{Mim}^{i0}[g^{FLRW}_{\mu\nu}] = V^{i0}_K[g^{FLRW}_{\mu\nu}] +\frac{ar^2\sin\theta}{\ka}\left(\ga^{ik}\dd_k\xi^0+\frac{1}{2}\Ga^i_{sn}\ga^{sn}\xi^0\right).
    \label{OriginalMimSuperpotFRLW}
\end{equation}
Generally speaking, last term in this formula is not zero even if one restrict $\xi^\mu$ to be constant and therefore superpotential \eqref{MimeticSuperpotential} cannot be reduced to Komar expression.\par

The impact of this issue on the meaningfulness of the superpotential \eqref{MimeticSuperpotential} mostly depends on whether new variables introduced by the change \eqref{MimeticChange} are observable on Einstein's equations. In this regard it is important to note, that there can be multiple $\sigma$ and $\gMim_{\al\be}$ for any given physical metric $g_{\mu\nu}$. Indeed, if one wants to find variables $\sigma$ and $\gMim_{\mu\nu}$ for the given metric, one has to solve differential equations. Equivalently, there is transformation group that maps solutions of these equations to itself. The exact form of all such transformations is not known (at least to the author's knowledge), but the similar problem occured in the mentioned Regge-Teitelboim gravity and was discussed
\footnote{
In the context of Regge-Teitelboim theory there exist many examples of how one GR solution can have multiple different embeddings, for examples see Schwarzshild \cite{statja27,kasner2, frons} or Reissner-N{\"o}rdstrom \cite{statja30} solutions.
} 
in many works
\cite{pogorelov, ivanova3, deser} (in work \cite{deser} it is also called "embedding gauges").
From \eqref{MimeticSuperpotential} it is obvious then that obtained superpotential seems to depend on the choice of $\sigma$ and hence depends on unobservable quantities for the solutions of Einstein's equations.
\par
This argument can also be applied to solutions for which Eistein's equations are satisfied only at some surface of constant time $\Sigma_t$. In this case $\sigma$ might be observable since Einstein's equations can be violated by the time evolution and hence in this case superpotential \eqref{MimeticSuperpotential} is acceptable. Though, the accurate analysis of this fact in the current framework involves canonical formalism, here we restrict ourselves to the simplistic proof why in this case $\sigma$ is still not physical. Equations of motion for mimetic gravity have the following form\cite{mukhanov}:
\begin{equation}
    G_{\mu\nu} + G\dd_\mu\sigma\dd_\nu\sigma = 0.
    \label{MimeticEOM}
\end{equation}
Let one of the GR equations $G^{00} = 0$ be satisfied at surface of constant time $\Sigma_t$. After substituting it into $00$-component of \eqref{MimeticEOM} and dividing by
$\dot{\sigma}$,
one immediately obtains $G = 0$, which leads to $G_{\mu\nu} = 0$ after using \eqref{MimeticEOM} again. Division here is possible because of the identity $g^{\mu\nu}\dd_\mu\sigma\dd_\nu\sigma = 1$, which follows directly from the change \eqref{MimeticChange}. 
Hence if $G^{00}$ is imposed in addition to \eqref{MimeticEOM}, Einstein's equations are satisfied too. One can use Bianchi identities to prove that time derivative of $G^{00}$ vanishes on Einstein's equations. Therefore, if one impose Einstein's equations at $\Sigma_t$ in addition to \eqref{MimeticEOM}, they will be satisfied afterwards.\par
Another theory under discussion - Regge-Teitelboim gravity - does also have pathological superpotential. This theory is described \cite{regge, davkar, pavsic85, deser} by the usual Einstein-Hilbert Lagrangian and transformation that presents physical metric $g_{\mu\nu}$ as induced metric of the $4$-dimensional manifold embedded into the flat bulk:
\begin{equation}
    g_{\mu\nu} = \eta_{ab}\dd_\mu y^a \dd_\nu y^b,
    \label{ReggeTeitelboimChange}
\end{equation}
where $\eta_{ab}$ is bulk metric, $y^a$ are embedding functions, and latin indices are related to the tensor representation of the $\eta_{ab}$ symmetry group. Since $y^a$ are scalars with respect to coordinate transformations, all the objects $Z^{a\rho\al_1..\al_k}$ in the superpotential \eqref{AntisymmetricSuperpotential} are symmetric over all of their space-time indices. However, it follows from \eqref{LGeneralExpr} that for the same reason the only source of upper indices in \eqref{LGeneralExpr} (and hence in \eqref{OmegaGeneralExpr}) are $Z^{a\rho\al_1..\al_k}$ for $k > 2$,
and given the previous argument they do not contribute
into the superpotential \eqref{AntisymmetricSuperpotential} due to its anti-symmetry.
Thus, it equals zero:
\begin{equation}
    V_{RT}^\rho = 0.
    \label{ReggeTeitelboimSuperpotential}
\end{equation}
This answer is obviously covariant, however, it carries nothing in terms of the physical information in contrast to Katz \cite{katz, julia} or Komar \cite{Komar, WaldBook} expressions. The most problematic point here is that \eqref{ReggeTeitelboimSuperpotential} doesn't pass any of the common test required for diffeomorphisms current such as correct energy for the isolated system  or correct energy to angular momentum ratio for Kerr solution (see chapter 6 in \cite{PetrovLompayTekinKopeikin} for example) and etc.\par
Discussed problems must be addressed if one wants Noether superpotential to reflect physical properties of the theory. One may always try to construct the superpotential defined for mimetic gravity specifically, that will satisfy all the mentioned requirements (like it was done numerous times before for many theories \cite{moller, BergmanConserv, BabakGrischukEnergy}). However, it is more desirable to somehow fix Noether procedure in order to solve these issues while preserving all known results computed before. As it will be shown in the next section, such way exists.

\section{Integrals of motion in theories with DFTA}\label{TheoriesWithChangeSection}
In the current section, we consider one class of the theories with higher derivatives in action - theories with DFTA. In general, one can describe this transformation by formula:
\begin{equation}
    \pp_A = \pp_A\left(\pp'_B,\dots,\dd_{\al_1..\al_W}\pp'_B\right).
    \label{GeneralChangeFormula}
\end{equation}
There are pretty old examples of such theories. For example, one may think of the usual metric formulation of General Relativity as the Palatini formulation if one take Levi-Civita connection formula as the change. However, only recently this way of modification of gravitational theories came to be seen as a perspective way to describe the dark matter phenomenon. The most important examples of such models were already mentioned multiple times. It is disformal gravity \cite{DisformalBekenstein, DisformalMimDeruelleRua} and especially its particular case - mimetic gravity, and Regge-Teitelboim embedding gravity \cite{regge, deser, davidson, rojas20,PastonNonRelativisticLimit, statja24, statja35}. It also was mentioned, that the main benefit one gets from this approach is that change adds some new degrees of freedom to the original theory while preserving all the original solutions. As it was shown in \cite{statja60, mukhanov2014}, this fact requires transformation to be differential, i.e. derivatives of the new independent variables must present in the change. These new degrees of freedom are of special interest because they can be used to described dark matter.\par
In the following we assume, that DFTA does not violate transformation law of the original fields. In particular, it means, that if there were a gauge symmetry in the original theory, it will stay untouched after the transformation was done. The main goal of the current section is to establish a relation between the conserved currents in theory before and after DFTA. As one will see, theories with DFTA have the additional relations for the coefficients $K_{(k)a}{}^{\rho\al_1\dots\al_{k-1}}$ which make this connection possible and greatly simplifies the analysis of the current in contrast to the general case. For brevity, we will continue to call everywhere theory before DFTA is applied {\it original} theory, and the theory with DFTA made we will call {\it new} theory.\par
In the next two subsection we will consider theories with no specific restrictions on the Lagrangians despite the condition $N_o \leq 2$ where $N_o$ denotes the order of derivatives in the action of the original theory. For any of these models we require transformation laws \eqref{fieldVarGeneralFormula} to obey $M = 1$ for both new and original fields. In particular, any tensor fields obviously satisfy these conditions. We also set the maximum order of the new fields in DFTA formula to 1 for these cases, i.e. $W = 1$. Such assumptions may seem to be too limiting, but they actually do not cross out any of the modern physically interesting theories, so further results can be safely applied to them. Though the cases with $M > 1$ will not be considered here, the author is certain, that all the techniques below can be successfully used there too. It is worth noting that the purpose of the subsection \ref{FirstOrderChangeSubsection} is to give very detailed discussion of all technicalities emerging in derivation of the current. Readers not interested in these details, can skip it and go straight to the subsection \ref{ScalarTensorChangeSubsection} for the results.\par
To reduce confusion, we will always define new theory by providing the lagrangian of the original theory and the formula for DFTA. Arguments of the transformation formula of the Lagrangian are assumed to be independent variables of new and original theories respectively (see \ref{Notations}).
\subsection{Case $N_o = 1$}
\label{FirstOrderChangeSubsection}
We start with the simplest non-trivial case of DFTA. Let the original theory be described by fields $\la_B$ and the Lagrangian:
\begin{equation}
    L = L(\la_B,\dd_\mu\la_B).
    \label{LambdaLagrangian}
\end{equation}
As it was in the section \ref{GeneralCascadeSection} we will assume, that the action has symmetry under some gauge group in the sense of the equality \eqref{ActionVariationGeneralCase}. The new theory is obtained from this one through the change:
\begin{equation}
    \la_B = \la_B\left(\pp_A,\dd_\mu\pp_A\right).
    \label{lambdaChange}
\end{equation}
One may also consider some background fields $\psi_C$ in addition to the dynamical ones $\la_B$. If these background fields have a non-trivial transformation law, they have to be taken into account in the r.h.s of the formula \eqref{NoetherTheoremStep1} and therefore in the current $J^\rho$ and also in $S^\rho$. We will return to this case later on. For now, we assume, that there are no background fields in both the Lagrangian \eqref{LambdaLagrangian} and the change \eqref{lambdaChange}.\par
The Lagrangian of the new theory obviously depends on the second derivatives of the fields $\pp_A$. To calculate the superpotential \eqref{AntisymmetricSuperpotential} and the current $J^\rho$ one should calculate the coefficients $L_{(p)a}{}^{\rho\al_1..\al_{p}}$ (they are defined in \eqref{JLDecomposition}). By using the definition \eqref{JDefinition} it is easy to derive the expressions for all the relevant coefficients:
\begin{align}
    &L^{(\pp)}_{(0)b}{}^\rho = Z^{A\rho}_{(\pp)}H^{(\pp)}_{(0)Ab}+Z^{A\rho\al}_{(\pp)}\dd_\al H^{(\pp)}_{(0)Ab}-\dd_\al Z^{A\al\rho}_{(\pp)} H^{(\pp)}_{(0)Ab},
    \label{L0GeneralFirstOrder}\\
    &L^{(\pp)}_{(1)b}{}^{\rho\be} = Z^{A\rho}_{(\pp)}H^{(\pp)}_{(1)Ab}{}^\be+Z^{A\rho\al}_{(\pp)}\dd_\al H^{(\pp)}_{(1)Ab}{}^\be+Z^{A\rho\be}_{(\pp)}H^{(\pp)}_{(0)Ab}-\dd_\al Z^{A\al\rho}_{(\pp)} H^{(\pp)}_{(1)Ab}{}^\be,\label{L1GeneralFirstOrder}\\
    &L^{(\pp)}_{(2)b}{}^{\rho\be\ga} = \frac{1}{2}\simup{Z_{(\pp)}^{A\rho\be}H^{(\pp)}_{(1)Ab}{}^\ga}{\be\ga}.\label{L2GeneralFirstOrder}
\end{align}
We also set $K^\mu = 0$ for simplicity. We will see further that this assumption does not add anything new in the final answer. Due to the field transformation, one can use chain rule to recalculate $Z_{(\pp)}^{A\be}$ and $Z_{(\pp)}^{A\rho\be}$ through the quantity $Z_{(\la)}^B$ from the original theory:
\begin{align}
    Z_{(\pp)}^{A\rho} = Z_{(\la)}^{B}\frac{\dd\la_B}{\dd\dd_\rho\pp_A} + Z_{(\la)}^{B\nu}\frac{\dd\dd_{\nu}\la_B}{\dd\dd_\rho\pp_A},
    \label{LambdaTheoryZFirstOrderThroughOldZ}\\
    Z_{(\pp)}^{A\rho\be} = Z_{(\la)}^{B\nu}\frac{\dd\dd_{\nu}\la_B}{\dd\dd_{\rho\be}\pp_A}.
    \label{LambdaTheoryZSecondOrderThroughOldZ}
\end{align}
Using these relations one may rewrite \eqref{L0GeneralFirstOrder}-\eqref{L2GeneralFirstOrder} in a more suitable form:
\begin{align}
\nonumber
    &L^{(\pp)}_{(0)b}{}^\rho = Z^{B}_{(\la)}\frac{\dd\la_B}{\dd\dd_\rho\pp_A}H^{(\pp)}_{(0)Ab}+Z^{B\nu}_{(\la)}\left[\frac{\dd\dd_{\nu}\la_B}{\dd\dd_\rho\pp_A}H_{(0)Ab}^{(\pp)}+\frac{\dd\dd_{\nu}\la_B}{\dd\dd_{\rho\al}\pp_A}\dd_\al H_{(0)Ab}^{(\pp)}-\dd_\al\left(\frac{\dd\dd_{\nu}\la_B}{\dd\dd_{\rho\al}\pp_A}\right)H_{(0)Ab}^{(\pp)}\right]-\\
    &\hspace{7cm}-\dd_\al Z^{B\nu}_{(\la)} \frac{\dd\dd_{\nu}\la_B}{\dd\dd_{\rho\al}\pp_A}H^{(\pp)}_{(0)Ab},
    \label{L0LambdaStep1}\\
    \nonumber
    &L^{(\pp)}_{(1)b}{}^{\rho\be} = Z_{(\la)}^{B}\frac{\dd\la_B}{\dd\dd_\rho\pp_A}H^{(\pp)}_{(1)Ab}{}^\be+Z_{(\la)}^{B\nu}\Bigg[\frac{\dd\dd_{\nu}\la_B}{\dd\dd_\rho\pp_A}H^{(\pp)}_{(1)Ab}{}^\be+\frac{\dd\dd_{\nu}\la_B}{\dd\dd_{\rho\al}\pp_A}\dd_\al H^{(\pp)}_{(1)Ab}{}^\be+\\
    &\hspace{3cm}+\frac{\dd\dd_{\nu}\la_B}{\dd\dd_{\rho\be}\pp_A}H^{(\pp)}_{(0)Ab}-\dd_\al\left(\frac{\dd\dd_{\nu}\la_B}{\dd\dd_{\al\rho}\pp_A}\right)H^{(\pp)}_{(1)Ab}{}^\be\Bigg]-\dd_\al Z_{(\la)}^{B\nu} \frac{\dd\dd_{\nu}\la_B}{\dd\dd_{\al\rho}\pp_A}H^{(\pp)}_{(1)Ab}{}^\be,
    \label{L1LambdaStep1}\\
    &L^{(\pp)}_{(2)b}{}^{\rho\be\ga} = \frac{1}{2}Z_{(\la)}^{B\nu}\simup{\frac{\dd\dd_{\nu}\la_B}{\dd\dd_{\rho\be}\pp_A}H^{(\pp)}_{(1)Ab}{}^\ga}{\be\ga}.
    \label{L2LambdaStep1}
\end{align}
For further analysis, it is convenient to use slightly different forms of these expressions, which can be obtained by applying Leibniz' rule to the last terms in the square brackets:
\begin{align}
\nonumber
    &L^{(\pp)}_{(0)b}{}^\rho = Z^{B}_{(\la)}\frac{\dd\la_B}{\dd\dd_\rho\pp_A}H^{(\pp)}_{(0)Ab}+Z^{B\nu}_{(\la)}\left[\frac{\dd\dd_{\nu}\la_B}{\dd\dd_\rho\pp_A}H_{(0)Ab}^{(\pp)}+2\frac{\dd\dd_{\nu}\la_B}{\dd\dd_{\rho\al}\pp_A}\dd_\al H_{(0)Ab}^{(\pp)}\right]-\\
    &\hspace{9cm}-\dd_\al \left(Z^{B\nu}_{(\la)} \frac{\dd\dd_{\nu}\la_B}{\dd\dd_{\rho\al}\pp_A}H^{(\pp)}_{(0)Ab}\right),
    \label{L0LambdaStep1AfterLeibniz}\\
\nonumber
    &L^{(\pp)}_{(1)b}{}^{\rho\be} = Z_{(\la)}^{B}\frac{\dd\la_B}{\dd\dd_\rho\pp_A}H^{(\pp)}_{(1)Ab}{}^\be+Z_{(\la)}^{B\nu}\Bigg[\frac{\dd\dd_{\nu}\la_B}{\dd\dd_\rho\pp_A}H^{(\pp)}_{(1)Ab}{}^\be+2\frac{\dd\dd_{\nu}\la_B}{\dd\dd_{\rho\al}\pp_A}\dd_\al H^{(\pp)}_{(1)Ab}{}^\be+\\
    &\hspace{6cm}+\frac{\dd\dd_{\nu}\la_B}{\dd\dd_{\rho\be}\pp_A}H^{(\pp)}_{(0)Ab}\Bigg]-\dd_\al \left(Z_{(\la)}^{B\nu} \frac{\dd\dd_{\nu}\la_B}{\dd\dd_{\al\rho}\pp_A}H^{(\pp)}_{(1)Ab}{}^\be\right).\label{L1LambdaStep1AfterLeibniz}
\end{align}
As we are aiming at relating the current in new theory $J^\rho_{(\pp)}$ to that one of the old $J^\rho_{(\lambda)}$, we should express $L^{(\pp)}_{(p)a}{}^{\rho\al_1..\al_{p}}$ through $L^{(\lambda)}_{(p)a}{}^{\rho\al_1..\al_{p}}$. From the formulae \eqref{L0LambdaStep1}-\eqref{L2LambdaStep1} it is clear, that this goal requires some additional identities to rewrite the coefficients for $Z^B_{(\la)}$ and $Z^{B\nu}_{(\la)}$. The source of these identities comes from the property of DFTA, assumed in preface of the section \ref{TheoriesWithChangeSection}. Namely, DFTA should preserve the transformation law of the original variables $\la_B$ with respect to the gauge group. This fact restricts the transformation law coefficients $H^{(\pp)}_{(k)Aa}{}^{\al_1..\al_k}$ of the new independent variables. Probably, the easiest way to obtain these restrictions is by expressing the infinitesimal variation of the $\la_B$ through the infinitesimal variations of $\pp_A$:
\begin{equation}
    \de\la_B = \frac{\dd\la_B}{\dd\pp_A}\de\pp_A+\frac{\dd\la_B}{\dd\dd_\rho\pp_A}\de\dd_\rho\pp_A.
\end{equation}
This equation holds for an arbitrary $\xi^a$. Hence, if one move all terms to the l.h.s, all the coefficients for $\xi^a$ and its derivatives must vanish independently:
\begin{align}
    & H_{(0)Bb}^{(\la)} = \frac{\dd\la_B}{\dd\pp_A}H_{(0)Ab}^{(\pp)}+\frac{\dd\la_B}{\dd\dd_\rho\pp_A}\dd_\rho H_{(0)Ab}^{(\pp)},\label{HZeroGeneralFirstOrderLambda}\\
    & H_{(1)Bb}^{(\la)}{}^\be = \frac{\dd\la_B}{\dd\pp_A}H_{(1)Ab}^{(\pp)}{}^\be+\frac{\dd\la_B}{\dd\dd_\be\pp_A} H_{(0)Ab}^{(\pp)}+\frac{\dd\la_B}{\dd\dd_\rho\pp_A} \dd_\rho H_{(1)Ab}^{(\pp)}{}^\be,
    \label{HFirstGeneralFirstOrderLambda}\\
    & 0 = \simup{\frac{\dd\la_B}{\dd\dd_\rho\pp_A}H^{(\pp)}_{(1)Ab}{}^\be}{\rho\be}.
    \label{LambdaWithRespectToDerPhiHOneIsAntisymmetric}
\end{align}
It is natural to use these restriction in \eqref{L0LambdaStep1} and \eqref{L1GeneralFirstOrder} to obtain contributions $L^{(\lambda)}_{(0)a}{}^{\rho}$ and $L^{(\lambda)}_{(1)a}{}^{\rho\be}$ (note, that for the original theory $L^{(\lambda)}_{(2)a}{}^{\rho\be\al} = 0$). However, in the current position one cannot do this, because \eqref{L0LambdaStep1}-\eqref{L2LambdaStep1} depend on objects like $\dd\dd_\nu\la_B/\dd\dd_\rho\pp_A$ and other derivatives from $\dd_\nu\la_B$.\par
Hopefully, these objects can be reduced to some linear combinations of derivatives of $\la_B$ with respect to the new fields $\pp_A$ and its derivatives. Since $\la_B$ depends on the coordinates only through the fields $\pp_A$, the following identity is satisfied:
\begin{equation}
    \dd_\nu\la_B = \frac{\dd\la_B}{\dd\pp_A}\left[\pp_A,\dd_\al\pp_A\right]\dd_\nu\pp_A+\frac{\dd\la_B}{\dd\dd_\rho\pp_A}\left[\pp_A,\dd_\al\pp_A\right]\dd_{\rho\nu}\pp_A.
    \label{LambdaChainRule}
\end{equation}
By applying operators $\dd/\dd\dd_{\al_1..\al_k}\pp_A$ to this identity and by properly choosing $k$ one obtains from it the following relations:
\begin{align}
    &\frac{\dd\dd_\nu\la_B}{\dd\pp_A} = \frac{\dd^2\la_B}{\dd\pp_{A'}\dd\pp_A}\dd_\nu\pp_{A'}+\frac{\dd^2\la_B}{\dd\dd_\rho\pp_{A'}\dd\pp_A}\dd_{\rho\nu}\pp_{A'},\label{lambdaDerWithRespectToPhi}\\
    &\frac{\dd\dd_\nu\la_B}{\dd\dd_\al\pp_A} = \frac{\dd^2\la_B}{\dd\pp_{A'}\dd\dd_\al\pp_A}\dd_\nu\pp_{A'}+\frac{\dd\la_B}{\dd\pp_A}\de^\al_\nu+\frac{\dd^2\la_B}{\dd\dd_\rho\pp_{A'}\dd\dd_\al\pp_{A'}}\dd_{\rho\nu}\pp_{A'},\label{lambdaDerWithRespectToDerPhi}\\
    &\frac{\dd\dd_\nu\la_B}{\dd\dd_{\al\ga}\pp_A} = \frac{1}{2}\simup{\frac{\dd\la_B}{\dd\dd_\al\pp_A}\de^\ga_\nu}{\al\ga}.\label{lambdaDerWithRespectToDerDerPhi}
\end{align}
On the other hand, one can use any other local function of the new filds $\pp_A$ instead of $\la_B$ in \eqref{LambdaChainRule}, for example, $\dd\la_B/\dd\pp_A$ or $\dd\la_B/\dd\dd_\rho\pp_A$. This path leads to another set of equations:
\begin{align}
    &\dd_\nu\left(\frac{\dd\la_B}{\dd\pp_A}\right) = \frac{\dd^2\la_B}{\dd\pp_{A'}\dd\pp_A}\dd_\nu\pp_{A'}+\frac{\dd^2\la_B}{\dd\dd_\eta\pp_{A'}\dd\pp_A}\dd_{\eta\nu}\pp_{A'},\label{PreCommutationRelations2}\\
    &\dd_\nu\left(\frac{\dd\la_B}{\dd\dd_\eta\pp_A}\right) = \frac{\dd^2\la_B}{\dd\pp_{A'}\dd\dd_\eta\pp_A}\dd_\nu\pp_{A'}+\frac{\dd^2\la_B}{\dd\dd_\al\pp_{A'}\dd\dd_\eta\pp_{A'}}\dd_{\al\nu}\pp_{A'}.
    \label{PreCommutationRelationsDer2}
\end{align}
By combining \eqref{PreCommutationRelations2},\eqref{PreCommutationRelationsDer2} with the expressions \eqref{lambdaDerWithRespectToPhi}, \eqref{lambdaDerWithRespectToDerPhi} it is easy to establish desired commutation relations:
\begin{align}
    &\left[\frac{\dd}{\dd\pp_A},\dd_\nu\right]\la_B = 0,
    \label{LambdaCommutator1}\\
    &\left[\frac{\dd}{\dd\dd_\al\pp_A},\dd_\nu\right]\la_B = \frac{\dd\la_B}{\dd\pp_A}\de_\nu^\al,
    \label{LambdaCommutator2}\\
    &\left[\frac{\dd}{\dd\dd_{\al\ga}\pp_A},\dd_\nu\right]\la_B = \frac{1}{2}\simup{\frac{\dd\la_B}{\dd\dd_\al\pp_A}\de^\ga_\nu}{\al\ga}.
    \label{LambdaCommutator3}
\end{align}
Through the use of these commutators it is now possible to link $L^{(\pp)}_{(0)a}{}^{\rho}, L^{(\pp)}_{(1)a}{}^{\rho\be}$, $L^{(\pp)}_{(2)a}{}^{\rho\be\al}$ with the coefficients, corresponding to the old current $J^\rho_{(\la)}$:
\begin{align}
    &L^{(\pp)}_{(0)b}{}^\rho = L^{(\la)}_{(0)b}{}^\rho+\frac{\de S}{\de \la_B}\frac{\dd\la_B}{\dd\dd_\rho\pp_A}H^{(\pp)}_{(0)Ab}+\frac{1}{2}\dd_\al \left(\antisimup{\frac{\dd\la_B}{\dd\dd_\rho\pp_A}Z^{B\al}_{(\la)}}{\rho\al}H^{(\pp)}_{(0)Ab}\right),\label{LambdaPhiTheoryL0Final}\\
    \nonumber
    &L^{(\pp)}_{(1)b}{}^{\rho\be} = L^{(\la)}_{(1)b}{}^{\rho\be}+\frac{1}{2}\Bigg[\frac{\de S}{\de\la_B}\frac{\dd\la_B}{\dd\dd_\rho\pp_A}H^{(\pp)}_{(1)Ab}{}^\be-\frac{\dd\la_B}{\dd\dd_\be\pp_A}H_{(0)Ab}^{(\pp)}Z^{B\rho}_{(\la)}\Bigg]^{\rho\be}+\\
    &\hspace{7cm}+\frac{1}{2}\dd_\al \left(\antisimup{\frac{\dd\la_B}{\dd\dd_\rho\pp_A}Z_{(\la)}^{B\al}}{\al\rho}H^{(\pp)}_{(1)Ab}{}^\be\right),\label{LambdaPhiTheoryL1Final}\\
    &L^{(\pp)}_{(2)\mu}{}^{\rho\be\al} = \frac{1}{4}\simup{\antisimup{\frac{\dd\la_B}{\dd\dd_\rho\pp_A}Z^{B\al}_{(\la)}}{\rho\al}H_{(1)Ab}^{(\pp)}{}^\be}{\al\be}.\label{LambdaPhiTheoryL2Final}
\end{align}
Finally, these expressions can be substituted in \eqref{JLDecomposition} to obtain the formula for the conserved current $J^\rho_{(\pp)}$ of the new theory (note sign in \eqref{fieldVarGeneralFormula} and also in \eqref{J'KDecomposition}):
\begin{equation}
    J^\rho_{(\pp)} = J^\rho_{(\la)}+\frac{\de S}{\de\la_B}\frac{\dd\la_B}{\dd\dd_\rho\pp_A}\de\pp_A+\frac{1}{2}\dd_\al\antisimup{Z^{B\al}_{(\la)}\frac{\dd\la_B}{\dd\dd_\rho\pp_A}\de\pp_A}{\al\rho}.
    \label{LambdaTheoryWithChangeCurrentFinal}
\end{equation}
The first two terms here were initially derived in \cite{ilinPastonUniverse}. The last one (in the mentioned work it was denoted as $I^\rho$), however, was just briefly discussed, and its explicit form was not found for the general case. Recall that at the beginning of this subsection, the term $K^\mu$ from the definition \eqref{JDefinition} was omitted. It is clear now, that for this quantity is just an additive term in the $J^\rho_{(\pp)}$, it will naturally fit the \eqref{LambdaTheoryWithChangeCurrentFinal} as part of $J^\rho_{(\la)}$, and the result \eqref{LambdaTheoryWithChangeCurrentFinal} still holds in this case.\par
The formula \eqref{LambdaTheoryWithChangeCurrentFinal} leads to some important conclusions:
\begin{enumerate}
    \item If DFTA \eqref{lambdaChange} does not contain any derivatives from $\pp_A$, then the old and new currents coincide.
    \item Let fields $\la_B,\pp_A$ be the tensors under gauge group transformation, the Lagrangian's arguments can be chosen in the form:
    \begin{equation}
        L = \hat{L}\left(\la_B,\D_\mu\la_B\right)
        \label{LambdaLagrWithCovariantDerivatives}
    \end{equation}
    and also $K^\mu$ is a vector density. We also introduce gauge connection $A_\mu{}^a{}_b$ into the theory. We assume, that it does not partake in DFTA. For the action does not depend on the derivatives of the connection which can be seen from \eqref{JDefinition}, its corresponding quantities $Z_{(A)}{}^\mu{}_a{}^b$ are not present in $J^\rho$. DFTA is assumed here to be the following:
    \begin{equation}
        \la_B = \hat{\la}_B\left(\pp_A,\D_\mu\pp_A\right).
    \end{equation}
    Note, that $\D_\mu$ in this formula depends on the connection $A_\mu{}^a{}_b$. It also may depend on any connection, for which the corresponding gauge symmetry obeys the condition $\de\la_B = 0$. The particular cases for such behavior are provided by Weyl invariance in the mimetic gravity \cite{mukhanov, MomeniNewModifiedMimetic} or the local Lorentz invariance in tetrad formulation of GR. None of the quantities $Z_{(\tilde{A})}, Z_{(A)}$ has non-trivial contribution in $J^\rho_{(\pp)}$, as there is no derivatives of these connections in DFTA.\par
    For described theory it can be easily shown, that $Z^{B\al}$ are covariant:
    \begin{equation}
        Z^{B\mu}_{(\la)}\equiv\frac{\dd L}{\dd\dd_\mu\la_B} = \frac{\dd \hat{L}}{\dd\D_\mu\la_B}.
        \label{lambdaTheoryZCovariantDerDerivative}
    \end{equation}
    The similar result holds for DFTA:
    \begin{equation}
        \frac{\dd\la_B}{\dd\dd_\mu\pp_A}=\frac{\dd\la_B}{\dd\D_\mu\pp_A}.
        \label{lambdaTheoryCovariantHigherJetDerivative}
    \end{equation}
    From here, one can immediately deduce that the second and third terms in \eqref{LambdaTheoryWithChangeCurrentFinal} are vector densities. Thus, for the case considered $J_{(\pp)}^\rho$ is also a vector density.
\end{enumerate}
From this points it is now clear that issues discussed in the section \ref{MimeticSuperpotentialSubsection} are not present in theories that are obtained from \eqref{LambdaLagrangian} by DFTA \eqref{lambdaChange}. However, the presented proof of covariance has the caveat. Consider case when the new or the old variables are constrained with conditions like $\D_\mu\la_B = f_{B\mu}$ or $\D_\mu\pp_B = f_{B\mu}(x)$, where $f_{B\mu}$ does not depend on $\la_B$ or $\pp_A$. In this situation one can no longer use the identities \eqref{LambdaLagrWithCovariantDerivatives} and \eqref{lambdaTheoryCovariantHigherJetDerivative}. Moreover, the Lagrangian cannot be reduced to the form \eqref{LambdaLagrWithCovariantDerivatives}. The possible workaround is to use gauge-invariant combinations as the arguments of the Lagrangian (and also the change) and then substitute the appropriate derivatives instead of \eqref{LambdaLagrWithCovariantDerivatives}. It is not clear how these gauge-invariant combinations can be chosen in current general setting, so this possibility will not be considered here.\par
Though the question about covariance is essential, there are no conditions in the case considered that ensure two currents equality for the solutions of the "old" theory, that are inevitably present in the new one. This problem was already encountered for specific theories in the section \ref{MimeticSuperpotentialSubsection}, but now it is apparent that it is present in the general case. We return to this issue later in the subsection \ref{AmbiguitySection}.\par
Before moving to the more general case, {there} are a few points that need clarification. First of all, until now, we assumed that there are no background fields in \eqref{LambdaLagrangian} and \eqref{lambdaChange}. If we allow them to appear in these expressions, the resulting answer for the current \eqref{LambdaTheoryWithChangeCurrentFinal} will eventually change. To derive the correct expression for the conserved current, one must treat these background fields $\pp_C$ in all the formulae in the full analogy with the dynamical ones while deriving the current. Apart from that, the logic that leads to \eqref{LambdaTheoryWithChangeCurrentFinal} does not change so the resulting current will be the following:
\begin{equation}
    J_{(\pp)}^\rho = J^\rho_{(\la)}+I^\rho[\text{independent variables}]+I^\rho[\text{background fields}],
    \label{JrhoWIthBackgroundFieldsFinal}
\end{equation}
where we introduced the notation:
\begin{equation}
    I^\rho[\pp_A] = \frac{\de S}{\de\la_B}\frac{\dd\la_B}{\dd\dd_\rho\pp_A}\de\pp_A+\frac{1}{2}\dd_\al\antisimup{Z^{B\al}_{(\la)}\frac{\dd\la_B}{\dd\dd_\rho\pp_A}\de\pp_A}{\al\rho}.
\end{equation}
\subsection{Case $N_o = 2$}\label{ScalarTensorChangeSubsection}
The procedure of relating $J^\rho_{(\pp)}$ to $J^\rho_{(\la)}$ for theory with the Lagrangian \eqref{LambdaLagrangian} and DFTA  
\eqref{lambdaChange} can be generalized to the more complex cases. At the moment, there is no known (at least, to the author's knowledge) generalization of the formula \eqref{LambdaTheoryWithChangeCurrentFinal} {for the arbitrary $N_o$. That being said, one can still generalize this to more interesting case $N_o = 2$:
\begin{equation}
    L = L(\la_B, \dd_\al \la_B,\dd_{\al\be}\la_B).
    \label{ScalarTensorLagrangian}
\end{equation}
and the DFTA \eqref{lambdaChange}}. The final answer for $J_{(\pp)}^\rho$ is presented below:
{
\begin{multline}
    J_{(\pp)}^\rho = J_{(\la)}^\rho+\frac{\de S}{\de \la_B}\frac{\dd \la_B}{\dd\dd_\rho\pp_A}\de\pp_A+\frac{1}{2}\dd_\be\antisimup{\left(Z_{(\la)}^{B\be}-2\dd_\ga Z_{(\la)}^{B\be\ga}\right)\frac{\dd\la_B}{\dd\dd_\rho\pp_A}\de\pp_A}{\rho\be}+\\+\frac{1}{3}\dd_{\be\ga}\antisimup{Z_{(\la)}^{B\ga\be}\frac{\dd\la_B}{\dd\dd_\rho\pp_A}\de\pp_A}{\rho\be}.
    \label{ScalarTensorChangeCurrentFinal}
\end{multline}
}The derivation of this result is not significantly different from the techniques used in the previous subsection. The only major difference{s} here are associated with the slightly more complex set of the commutations relations \eqref{LambdaCommutator1},\eqref{LambdaCommutator2} и \eqref{LambdaCommutator3}. The explicit formulae for them can be found in the Appendix \ref{ScalarTensorCommutationRelationsAppendix}\par
As it was for \eqref{lambdaChange}, if transformation \eqref{lambdaChange} does not contain the derivatives, conserved currents in the old and new theories coincide. However, in the general case, the current \eqref{ScalarTensorChangeCurrentFinal} does not reduce to the original one when the "old" equations of motion are satisfied $\de S/\de\la_B = 0$ as one can expect. In the next subsection, we will discuss this problem in more detail and propose a possible solution.

\subsection{Ambiguity in Noether's procedure. Covariantization of the currents}
\label{AmbiguitySection}
Despite the notable popularity of the procedure \eqref{NoetherTheoremStep1}-\eqref{OffShellTotalNoetherIdentity} to construct the conserved currents, in some situations it may not give a unique answer. This ambiguity is tied to the high order derivatives of the independent variables in the action ($N\geq2$) and can be described as follows. Consider once again the formula \eqref{NoetherTheoremStep1}. By definition, the quantities $Z^{A\al_1\dots\al_i}$ are symmetric over the indices $\al_1,\dots,\al_i$. It is easy to see that this formula does not change under the transformation:
\begin{equation}
    \tilde{Z}^{A\al_1\dots\al_i}= Z^{A\al_1\dots\al_i}+N^{A\al_1\dots\al_i},
    \label{AmbiguityTransform}
\end{equation}
where $N^{A\al_1\dots\al_i}$ has zero symmetric part with respect to all indices $\al_1\dots\al_i$ excluding those from multi-index $A$. However, the consequent use of Leibniz' rule, which is needed to proceed to the $J^\rho$, can in general lead to the different answers for the first term in \eqref{JDefinition}. Indeed, there are choices of \eqref{AmbiguityTransform}, which lead to $\tilde{Z}^{A\rho\al_1\dots\al_{i-1}}$, that will no longer be symmetric with respect to $\rho$ and any other index. This altered procedure does not change the r.h.s of the \eqref{OffShellTotalNoetherIdentity}, so any two currents corresponding to the different choices $N^{A\al_1\dots\al_i}$ can differ only by a divergence of the superpotential:
\begin{equation}
    \tilde{J}^\rho-J^\rho = \dd_\be W^{\be\rho}.
    \label{AmbiguityInCurrents}
\end{equation}
This is the ambiguity which was mentioned above \footnote{In work \cite{PetrovCovariantized} the ambiguity \eqref{AmbiguityTransform} is parametrized in another way, namely, by the choice of the arguments in the Lagrangian. At the moment, it is not clear enough how this method and the transformation \eqref{AmbiguityTransform} agree with each other.}.\par
Some points about this property should be clarified. First of all, it is much smaller than one, which arises from using \eqref{OffShellTotalNoetherIdentity} as a definition for the currents $J^\rho$ (see, for example, \cite{BarnichGeneralForm}), and only appears in the procedure \eqref{NoetherTheoremStep1}-\eqref{JDefinition}. Secondly, the formula \eqref{AmbiguityTransform} can be used not only for the cases of the gauge symmetries but also for cases with just global ones. Thirdly, recall that for the definition of the conserved charges, one needs to ensure the l.h.s of the \eqref{JFormStokesTheorem} equals zero. These conditions are fulfilled by restricting the class of functions for which one seeks the action minimum. It was stated, that one may take conditions \eqref{KAsymptoticCondition}, \eqref{ActionPrincipleConsistency} for this purpose. The ambiguity \eqref{AmbiguityTransform} affects these restrictions because the quantity $\al^\mu$ actually coincides with the first term in \eqref{JDefinition} and also is derived similarly. However, it is obvious, that the transformation \eqref{AmbiguityTransform} just shifts $\al^\mu$ by the divergence of the superpotential:
\begin{equation}
    \al^\mu\rightarrow\al^\mu+\dd_\be h^{\be\mu},\;\;\; h^{\be\mu} = - h^{\mu\be}.
\end{equation}
By using the Stoke's theorem and fixing time derivatives of $\pp_A$ up to the order $N-1$ on $\Sigma_1$ and $\Sigma_2$ it can be easily seen, that the relation \eqref{ActionPrincipleConsistency} does not change.\par
{It is important to understand that even after fixing $N^{A\al_1\dots\al_i}$ there is still possibility to obtain different results from the Noether procedure. Indeed, once $\tilde{Z}^{A\rho\al_1\dots\al_{i}}$ are not symmetric over their indices after $A$, one may differently choose the order of derivatives on $\pp_A$ when applying Leibniz rule in \eqref{Leibniz'Rule}. The result \eqref{OffShellTotalNoetherIdentity} is obviosuly will not change unlike the current $J^\rho$. In order to totally fix the ambiguity in the Noether's procedure one must also define the order of applying Leibniz' rule to expansion \eqref{NoetherTheoremStep1}. By definition we fix the order of application of Leibniz' rule in the way it leads to the following expression for $J^\rho$:
\begin{equation}
    J^{\rho} \equiv \sum_{j = 1}^N\sum_{i = 0}^{j-1}(-1)^{i+j+1}\dd_{\al_1\dots\al_{j-i-1}}\tilde{Z}^{A\al_1\dots\al_{j-i-1}\rho\al_{j-i}\dots\al_{j-1}}\dd_{\al_{j-i}\dots\al_{j-1}}\de\pp_A+K^\rho.
    \label{JAmbiguityFixedDefinition}
\end{equation}
Until the end of the paper this formula will be used for the calculation of the current $J^{\rho}$. It is easy to check that this definition coincide with the standard one from the textbooks (see, for example, \cite{PetrovLompayTekinKopeikin}) in case of $N = 1$.
}\par
Fixation of the ambiguity can be potential tool for fixing the issues with the currents, obtained in the previous subsection. They were mentioned before multiple times. The first one is the violation of new and old currents equatlity for the "old" solutions:
\begin{equation}
    J_{(new)}^\rho[\psi_A] \neq J_{(original)}^\rho[\psi_A].
    \label{JNewIsEqualToJOriginal}
\end{equation}
The second issue is the violation of covariance: the terms arising in the expression for the conserved current can break vector density transformation law, which prevents one to correctly define the charges \eqref{ChargeDef}. These problems can be solved, for example, by simply omitting undesirable terms with the redefinition of currents because they are divergences of the superpotentials, see \eqref{LambdaTheoryWithChangeCurrentFinal} and \eqref{ScalarTensorChangeCurrentFinal}. As it was noted at the end of subsection \ref{MimeticSuperpotentialSubsection}, a more preferable opportunity is to fix the ambiguity \eqref{AmbiguityTransform} in Noether's procedure in the way the resulting current will be free of the mentioned problems. Here we show, that for theories from the subsections \ref{FirstOrderChangeSubsection} and \ref{ScalarTensorChangeSubsection} it can be done by proper ambiguity fixation.\par
We start with the theory described by the Lagrangian \eqref{LambdaLagrangian} and DFTA \eqref{lambdaChange}. It is obvious that in this theory the only source of ambiguity is transformation \eqref{AmbiguityTransform} for $Z_{(\pp)}^{A\rho\al}$. By using commutator \eqref{LambdaCommutator3} we can write this object like the following:
\begin{equation}
    Z_{(\pp)}^{A\rho\al} = \frac{1}{2}\simup{Z_{(\la)}^{B\rho}\frac{\dd\la_B}{\dd\dd_\al\pp_A}}{\rho\al}.
\end{equation}
The key advantage of this formula is that the symmetry of the r.h.s is explicitly controlled by the symmetrizer. We can then choose $N^{A\rho\be}$ in the form:
\begin{equation}
    N^{A\rho\be}\equiv\frac{1}{2}\antisimup{Z_{(\la)}^{B\rho}\frac{\dd\la_B}{\dd\dd_\al\pp_A}}{\rho\al}
\end{equation}
which leads to the following $\tilde{Z}_{(\pp)}^{A\rho\al}$:
\begin{equation}
    \tilde{Z}_{(\pp)}^{A\rho\al}\equiv Z_{(\la)}^{B\rho}\frac{\dd\la_B}{\dd\dd_\al\pp_A}.
    \label{LambdaTheoryTildeZChoice}
\end{equation}
In order to obtain the formula similar to \eqref{LambdaTheoryWithChangeCurrentFinal} one should substitute this expression in the $L^{(\pp)}_{(0)Aa}{}^{\rho}$, $L^{(\pp)}_{(1)Aa}{}^{\rho\be}$ and $L^{(\pp)}_{(2)Aa}{}^{\rho\be\ga}$. However, it will be a mistake to use the standard formulae \eqref{L0GeneralFirstOrder}-\eqref{L2GeneralFirstOrder} for them, because they were derived under the assumption of the index symmetry of $\tilde{Z}^{A\rho\al_1\dots\al_k}$, which is not anymore the case. None the less, it is easy to check, that \eqref{L0GeneralFirstOrder}-\eqref{L2GeneralFirstOrder} {agrees with the definition \eqref{JAmbiguityFixedDefinition}} even without index symmetries {of} $\tilde{Z}^{A\rho\al_1\dots\al_k}$. By passing them into \eqref{LambdaTheoryWithChangeCurrentFinal} and following the logic of the subsection \ref{FirstOrderChangeSubsection}, we can finally obtain simple formula for the current:
\begin{equation}
    \tilde{J}^\rho = J^\rho_{(\la)}+\frac{\de S}{\de\la_B}\frac{\dd\la_B}{\dd\dd_\rho\pp_A}\de\pp_A.
    \label{LambdaTheoryAmbiguityUsedCurrent}
\end{equation}
This expression for the current is much more convenient and does not posses {\it both} issues mentioned at the start of the subsection. Indeed, it immediately follows from the discussion at the end of the subsection \ref{FirstOrderChangeSubsection}, that the current \eqref{LambdaTheoryAmbiguityUsedCurrent} is a vector density if $\la_B, \pp_A$ transforms as a tensors under the gauge transformations, and the action is gauge-invariant, i.e. $K^\mu = 0$ in \eqref{ActionVariationGeneralCase}. It is also obvious, that it coincides with the original theory current $J^\rho_{(\la)}$ on the solutions of the original theory.\par
From here one may be interested in the decomposition \eqref{SVCurrentDecomposition} for the current \eqref{LambdaTheoryAmbiguityUsedCurrent}. After some calculations, we obtain the result:
\begin{equation}
     \tilde{J}^\rho = S_{(\pp)}^\rho+\dd_\be\left(V^{\rho\be}_{(\la)}+\frac{1}{2}\frac{\de S}{\de\la_B}\antisimup{\frac{\dd\la_B}{\dd\dd_\be\pp_A}H^{(\pp)}_{(1)Ab}{}^{\rho}}{\rho\be}\xi^b\right),
     \label{LambdaTheoryCurrentCorrectedFinalSV}
\end{equation}
where $V^{\rho\be}_{(\la)}$ is a superpotential of the original theory. If the arguments in \eqref{lambdaChange} can be chosen in a gauge-invariant way, and the fields $\la_B, \pp_A$ transforms as tensors under the gauge transformation, the superpotential from \eqref{LambdaTheoryCurrentCorrectedFinalSV} will be a tensor density with respect to this group and hence is valid for the definition of the charges \eqref{ChargeDef}.\par
Similarly to the previous discussion, one may derive "corrected" current {for theories of the type \eqref{ScalarTensorLagrangian}} with DFTA \eqref{lambdaChange}. It can be shown, that there exist such choice of {$N_{(\pp)}^{A\rho\be}, N_{(\pp)}^{A\rho\be\ga}$, that the objects $Z_{(\pp)}^{A\al\be}$,  $Z_{(\pp)}^{A\al\be\ga}$} will take the form:
{
\begin{align}
    &\tilde{Z}_{(\pp)}^{A\rho\be\al}\equiv Z_{(\la)}^{B\rho\be}\frac{\dd \la_B}{\dd\dd_\al\pp_A},\label{ScalarTensorTildeZ3SigmaChoice}\\
    &\tilde{Z}_{(\pp)}^{A\rho\be} = Z_{(\la)}^{B\rho}\frac{\dd \la_B}{\dd\dd_\be\pp_A}+Z_{(\la)}^{B\rho\be}\frac{\dd\la_B}{\dd\pp_A}+2Z_{(\la)}^{B\rho\nu}\dd_\nu\left(\frac{\dd \la_B}{\dd\dd_\be\pp_A}\right).
    \label{ScalarTensorTildeZ2SigmaChoice}
\end{align}
For this choice of $\tilde{Z}$ after some calculations one can show that the corresponding current has the form \eqref{LambdaTheoryCurrentCorrectedFinalSV}. Interestingly enough, there are no new contributions into the superpotential. It seems from the above formulae that the expression for the "corrected" superpotential does not depend on $N_o$. In the subsection \ref{LagrangeMultiplierApproachSection} a possible explanation of this behavior is given.}\par
As it was mentioned, the ambiguity appears only in theories with high-order derivatives in the action. Hence in the case considered the Noether's procedure in the original theory is also ambiguous. It is worth noting, that the transformation \eqref{AmbiguityTransform} for $Z^{B\al\be}_{(\la)}$ in the original theory cannot be reduced to that one for $\tilde{Z}_{(\pp)}^{A\al\be}$ in the new theory. Therefore, one needs to specify the choice of the quantities $N$ for the original {quantity} $Z^{A\al\be}_{(\la)}$ in the formulae \eqref{ScalarTensorTildeZ3SigmaChoice},\eqref{ScalarTensorTildeZ2SigmaChoice}. It is obvious, that in these relations $Z^{A\al\be}_{(\pp)}$ must be symmetric with respect to the pair $\al\be$ because these formulae originate from the chain rule.

\section{Applications}\label{Applications}
\subsection{Changes without derivatives}
In the previous section we proved, that for theory \eqref{ScalarTensorLagrangian} with the change \eqref{lambdaChange} the corresponding original and new currents coincide for both the original Noether's procedure \eqref{ScalarTensorChangeCurrentFinal} and the corrected one \eqref{LambdaTheoryCurrentCorrectedFinalSV}. It allows one to easily calculate Noether, currents if the change does not contain derivatives. For example, it follows from \eqref{LambdaTheoryCurrentCorrectedFinalSV}, that for the Weyl transformation with scale factor that doesn't contain field derivatives and for the tetrad change $g_{\mu\nu} = e_\mu^ae_{a\nu}$ the corresponding currents do not change at all regardless of the "old" theory. For example, this transformation allows one to switch between Jordan and Einstein frames in F(R) gravity \cite{BahamondeOdintsovJordanF(R)Corresp2016,SotiriouFaraoniF(R)GravityRev}. It should be noted though, that the term $S^\rho$ in decomposition \eqref{SVCurrentDecomposition} may still change, as the equations of motion may look different.

\subsection{Corrected superpotentials for mimetic gravity and Regge-Teitelboim theory}\label{CorrectedSuperpotsSubsection}
{Let us return to the cases considered in the subsection \ref{MimeticSuperpotentialSubsection}: mimetic gravity and Regge-Teitelboim theory. In these theories the original action is usually taken as standard Einstein-Hilbert action, so $N_o = 2$. Since DFTA \eqref{MimeticChange} and \eqref{ReggeTeitelboimChange} don't contain any derivatives of the independent variables with $M \geq 1$, the last term in the superpotential \eqref{LambdaTheoryCurrentCorrectedFinalSV} vanishes. Thus, in both cases the corrected superpotentials reduce to the Komar 2-form:}
\begin{equation}
    V^{\be\rho}_K = \frac{\sqrt{-g}}{2\ka}\antisimup{\D^\be\xi^\rho}{\rho\be}.
    \label{CorrectedMimeticSuperpotential}
\end{equation}
This statement holds for the mentioned mimetic, disformal, and also Regge-Teitelboim theories of gravity. From here one immediately concludes that for all these theories diffeomophisms-related integrals of motion reproduce all results for integrals of motion for GR such as mass\cite{Komar}. It is interesting that generally the last term in \eqref{LambdaTheoryCurrentCorrectedFinalSV} is not zero, however, it vanishes every time transformation \eqref{GeneralChangeFormula} does not contain new variables with $M \geq 1, W \geq 0$:
\begin{equation}
    J_{(\pp)}^\rho = S^\rho_{(\pp)} + \dd_\al V^{\al\rho}_{K},
    \label{simplifiedJ}
\end{equation}
At the moment, the physics behind this last term in \eqref{LambdaTheoryCurrentCorrectedFinalSV} is not clear, as in the case considered the corresponding contribution to the superpotential simply vanished. However, it is not true in more complex cases. For example, one may consider modification of the initial idea of mimetic gravity with the change involving vectors $A_\mu$ instead of scalars \cite{F2AbelianMimeticGorjiMukhoyama}, \cite{F2YangMillsMimeticGorjiMukhoyama}:
\begin{equation}
    g_{\mu\nu} = \gMim_{\mu\nu}\sqrt{\be\gMim^{\al\be}\gMim^{\ga\de}F_{\al\ga}F_{\be\de}},
    \label{F2MimeticChange}
\end{equation}
where $\be \equiv \pm 1$, and $F_{\mu\nu}$ is just field strength for $A_\mu$. One may also  For $A_\mu$ one has transformation law with $M = 1$, so the formula \eqref{LambdaTheoryCurrentCorrectedFinalSV} leads to the following result for the current:
\begin{equation}
    J_{(A)}^\rho = S^\rho_{(A)} + \dd_\al \left(V^{\al\rho}_K+\frac{1}{\ka}\sqrt{-g}G F^{\be\rho}A_\ga\xi^\ga\right).
    \label{F2MimeticSuperpotential}
\end{equation}
The second term in the superpotential is exactly the contribution from the new term in \eqref{LambdaTheoryCurrentCorrectedFinalSV}. This term, however, is not invariant under $U(1)$-transformations, so one should somehow address this in order to properly define integrals of motion. It can be easily checked, that this term coincides (modulo factor $G$) with electromagnetic field contribution in the superpotential for Einstein-Maxwell theory. Thus, one may apply some other known methods of covariantization of E-M theory to this case (by, for example, fixing gauge depending on $\xi^\mu$ \cite{linden-bell}).\par
{It is worth noting, that the formula \eqref{LambdaTheoryCurrentCorrectedFinalSV} does not bring into superpotential any dependencies on the new fields like $\bar{g}_{\mu\nu}$ or $\sigma$ for mimetic/disformal gravity or embedding functions for $y^a$ in the Regge-Teitelboim gravity. In the context of the first one it is not big surprise since mimetic dark matter is actually a pressureless dust \cite{mukhanov}. However, for Regge-Teitelboim gravity formula \eqref{CorrectedMimeticSuperpotential} ensures that the extra matter contributes to the energy in the similar way as the usual matter, as no additional contributions to the total superpotential arise. Formula \eqref{simplifiedJ} makes one sure that this statement is true for any theory with DFTA satisfying $M \geq 1, W > 0$.}\par
{There is another problem with the current for the Regge-Teitelboim theory that was not discussed before. Namely, for this theory (and for any other theory with fields for which $M = 0$) one can easily check that $S^\rho = 0$ , which immediately follows from \eqref{SDef}. This result is obviously not particularly tied to Regge-Teitelboim theory - it is valid for any theory with fields that only have $M = 0$ (see \cite{ilinPastonUniverse}). This issue remains unresolved, and, as one can see from \eqref{LambdaTheoryCurrentCorrectedFinalSV}, fixation of the ambiguity \eqref{ScalarTensorTildeZ3SigmaChoice}, \eqref{ScalarTensorTildeZ2SigmaChoice} is not useful in this situation at all. This is interesting topic for the future investigation, however, in the current article we are not going into more details about it.}\par

\subsection{Mass, Angular momentum and first law of BH thermodynamics}\label{MassEntropySection}
As it was discussed in the section \ref{ChargesSubsection}, the superpotential allows one to easily define conserved asymptotic charges. In case of diffeomorphisms $J^\rho$ depends on the infinitesimal vector $\xi^\mu$, so one needs to specify it to define physically meaningful conserved quantities. It is natural to use asymptotically Killing vectors for this purpose (see, for example, \cite{Komar, katz, DeruelleKatzConformalMass}). For asymptotically flat spacetime one may choose 10 such vectors, which upon substitution in the formula \eqref{ChargeDef} will give rise to the total energy-momentum and angular momentum of the spacetime.\par
As was stated in the previous subsection, if the change doesn't contain fields with $M\geq 1, W > 0$, then the corresponding superpotentials (and therefore integrals of motion) will not change. For the cases considered above (except vector generalization of mimetic gravity) the corrected superpotential \eqref{LambdaTheoryCurrentCorrectedFinalSV} leads to the same Komar energy and angular momentum (see \eqref{CorrectedMimeticSuperpotential}) as they are for GR.
It is well-known that these specific integrals leads to the incorrect angular momentum to the mass ratio in the Kerr solution \cite{katz, julia} {and also to only half of the ADM mass (see, for example, \cite{DeruelleKatzConformalMass})}. That's why usually one starts not with the Einstein-Hilbert action but rather with first-order derivative action for the metric GR. It was shown in \cite{katz}, that this action leads to the so-called Katz superpotential. It is obvious that this case {still fits} the symmetry {definition} \eqref{ActionVariationGeneralCase}, and hence the formula \eqref{LambdaTheoryCurrentCorrectedFinalSV} can still be used to construct the conserved currents. However, the action of the theory is no longer generally covariant, so the resulting current may not be a vector density and needs to be covariantized in the standard way \cite{katz}. For more complex cases one can still use many other covariantization techniques, for example, \cite{linden-bell}. If one considers the change \eqref{F2MimeticChange}, there will be another contribution from the second term in \eqref{F2MimeticSuperpotential} into all integrals of motion. However, as this term is not gauge-invariant under $U(1)$ rotations, one needs to deal with this term by either choosing specific boundary conditions that are consistent with \eqref{ActionPrincipleConsistency}, or by modifying the Noether's procedure further than it is presented here. \par
One important application of the Noether currents \eqref{JDefinition} is their direct connection to the first law of black hole thermodynamics. We consider a black hole solution with bifurcate Killing horizon in theory \eqref{ScalarTensorLagrangian} with the change \eqref{lambdaChange}. To obtain first law from Noether's current \eqref{JDefinition} we are following here Wald's formalism proposed in \cite{WaldEntropy, IyerWaldEntropy}. It was established in \cite{IyerWaldEntropy}, that the entropy for BH is given by formula:
\begin{equation}
    S = 2\pi \int_\mathcal{B}X^{\mu\nu}\ep_{\mu\nu},
    \label{IWEntropy}
\end{equation}
where $\mathcal{B}$ is bifurcation $2$-surface, and $\ep_{\mu\nu}$ is binormal to $\mathcal{B}$, and $X^{\mu\nu}$ is coefficient in the expansion of the current $J^\rho$ (in differential form notation):
\begin{equation}
    V = W_\mu\xi^\mu+X^{\mu\nu}\D_{[\mu}\xi_{\nu]}+Y+dZ.
    \label{WaldEntropyDecomposition}
\end{equation}
It was also proven in work \cite{IyerWaldEntropy}, that this expansion for the current is always possible. Comparing this expansion to the formula \eqref{LambdaTheoryCurrentCorrectedFinalSV} it is clear, that the new term in superpotential that arise from the change of variables does not contain any derivatives of $\xi^\mu$. Hence it doesn't contribute to the form $X^\mu\nu$, which leads to the simple result for the Iyer-Wald entropy $S_{(\pp)}$ in theory after change:
\begin{equation}
    S_{(\pp)} = 2\pi\int_\mathcal{B}X_{(\la)}^{\mu\nu}\ep_{\mu\nu},
\end{equation}
where $X_{(\la)}^{\mu\nu}$ is coefficient in decomposition \eqref{WaldEntropyDecomposition} of the current $J^\rho_{(\la)}$, i.e. the original current. It should be stressed that this formula is valid for any asymptotically flat BH with bifurcate Killing horizon, not just for the "shadow" BH solutions, that are inherited from the original theory. From the recent perspective, however, it should be noted, that extra BH solutions in theories with change are usually non-asymptotically flat (see, for example, \cite{GorjiMimeticBlackHoles}), and one must account that in the following calculations. Special attention must also be paid to the changes with additional gauge symmetries like \eqref{F2MimeticChange}. In these situations first law will have the additional contribution arising from the corresponding gauge charges. The exact computations for specific black holes in above mentioned theories will be performed elsewhere.\par

\subsection{Cosmology}\label{CosmologyAppSection}
Another useful application of the formulae \eqref{LambdaTheoryCurrentCorrectedFinalSV} and \eqref{LambdaTheoryAmbiguityUsedCurrent} is the derivation of the integral constraints for the localized perturbations on FLRW background. In works \cite{Traschen,TraschenErdley} such constraints were used to significantly reduce Sachs-Wolfe effect on the angular CMB fluctuations for General Relativity. To re-derive these constraints in other theories one may follow the appoach proposed in \cite{linden-bell}. We will not dive into the details (as the results highly depend on the specific changes of variables) and restrict ourselves to giving qualitative arguments.\par
Consider theory with the Lagrangian \eqref{ScalarTensorLagrangian} with metric $g_{\mu\nu}$ components being the original fields $\la_B$. The corresponding change of variables \eqref{lambdaChange} has the form:
\begin{equation}
    g_{\mu\nu} = g_{\mu\nu}(\pp_A, \dd_\mu\pp_A),
    \label{cosmChange}
\end{equation}
We introduce two distinct spacetimes - physical $\mathcal{M}$ and background $\bar{\mathcal{M}}$ - with metrics $g_{\mu\nu}$, $\bar{g}_{\mu\nu}$ with corresponding change fiels $\pp_A, \bar{\pp}_A$ respectively. We assume that $\mathcal{M}$ and $\bar{\mathcal{M}}$ can be mapped one to each other, so one can introduce the same coordinates on both manifolds through this mapping. It also means, that if one performs a coordinate transformation, it affects both metrics and therefore change fields defined on $\mathcal{M}$ and $\bar{\mathcal{M}}$. Such constructions were used in the old approach of Rosen \cite{RosenBimetric}.\par
Though all the abovementioned theories are based on the Einstein-Hilbert action, it is more convenient to use action with the standard $"\Ga\Ga-\Ga\Ga"$ Lagrangian with subtracted fully background term:
\begin{equation}
    L_{gr} = L_{\Ga\Ga}-\bar{L}_{\Ga\Ga}, \;\;\; S_{gr}\equiv \int_\mathcal{M} d^Dx L_{gr}.
    \label{grLagrBackground}
\end{equation}
where $\bar{L}_{\Ga\Ga}$ is the same as $L_{\Ga\Ga}$ but depends on the background fields only. We also assume, that there are additional matter Lagrangian in the total action $S$.
This construction obviously doesn't change field equations of GR, but Noether current that corresponds to diffeomorphisms will have the additional contribution from $\bar{L}_{\Ga\Ga}$, as background fields also transform under diffeomorphisms and thus contribute to the current. For implications we will be interested in "strongly" conserved currents $J'^\rho$ that was introduced in \eqref{J'Definition}, \eqref{J'ConsLaw}. We denote this strong current for the Lagrangian \eqref{grLagrBackground} as $I^\rho$. To proceed further it is useful to write special decomposition for this current as it was done in \cite{linden-bell}:
\begin{equation}
    I_{gr}^\rho = \theta^\mu{}_\nu\xi^\nu+\sigma^{\rho[\mu\nu]}\D_{[\mu}\xi_{\nu]}+\zeta^\mu(\xi^\ga, \dd_\al\xi^\ga, \dd_{\al\be}\xi^\ga).
    \label{RelativeBackgroundCurrent}
\end{equation}
From definition \eqref{J'Definition} it follows, that $\theta^\mu{}_\nu$ contains contribution from $X^\rho$. As it will be noted later, this quantity plays central role in the approach as it leads to the perturbed energy-momentum tensor factor in the integral constraints. For theories with $M \leq 1$ it has the simple form:
\begin{equation}
    X_{gr}^\rho = -\frac{\de S_{gr}}{\de \pp_A}H_{(1)A\mu}{}^\rho\xi^\mu.
\end{equation}
Assuming that the equations of motion for the total action are satisfied, one can express this contribution through symmetric energy-momentum tensor:
\begin{equation}
    X_{gr}^\rho = -\frac{1}{2}\left(\sqrt{-g}T^{\mu\nu}\frac{\dd g_{\mu\nu}}{\dd\pp_A}-\dd_\al\left(\sqrt{-g}T^{\mu\nu}\frac{\dd g_{\mu\nu}}{\dd\dd_\al\pp_A}\right)\right)H_{(1)A\mu}{}^\rho\xi^\mu.
    \label{XAfterChange}
\end{equation}
This expression is generalization of the first term in the formula (2.17) in \cite{linden-bell}. In case of mimetic gravity the only new field that will contribute to \eqref{XAfterChange} is auxiliary metric, so it is clear from the change \eqref{MimeticChange}, that $X^\rho$ will be equal to $-\frac{1}{2}\sqrt{-g}T^\mu_\nu\xi_\nu$. However, in more complex cases both terms can give non-trivial contribution. For example, in case of gauge mimetic gravity \eqref{F2MimeticChange} the contribution from terms with $\pp_A = \gMim_{\mu\nu}$ will be the same as in the usual mimetic gravity while the contribution from terms with $\pp_A = A_\mu$ have zero first term and non-zero second one.\par
Once $X_{gr}^\rho$ and other coefficients are calculated, the path to integral constraints is straightforward. Consider the localized perturbations of the change fields $\pp_A$, so that the perturbed physical metric $g_{\mu\nu}$ will have FLRW symmetry. For the background one should pick de-Sitter spacetime. Mapping between perturbed and background spaces can be done by firstly introducing 3+1 decomposition in both spaces and then by mapping constant time hypersurfaces from $\mathcal{M}$ and $\bar{\mathcal{M}}$ to each other. Due to the properties of de-Sitter spacetime such mapping is always possible for any topology of spacelike slices in $\mathcal{M}$. It is clear from the definition \eqref{J'Definition} and decomposition \eqref{SVCurrentDecomposition}, that $I^\rho = \dd_\al \left(V^{\al\rho}-\bar{V}^{\al\rho}\right)$, where $V^{\al\rho}$ is defined in \eqref{AntisymmetricSuperpotential}. One may then choose slice $\Sigma$ from $3+1$ decomposition of $\mathcal{M}$ and integrate this identity over it:
\begin{equation}
    \int_{\Sigma} d\Sigma_\mu\left(\theta^\mu{}_\nu\xi^\nu+\sigma^{\rho[\mu\nu]}\D_{[\mu}\xi_{\nu]}+\zeta^\mu\right) = \int_{\dd\Sigma} d\Sigma_{\mu\nu}\left(V^{\al\rho}-\bar{V}^{\al\rho}\right).
    \label{TraschenLikeConstraints}
\end{equation}
For GR with $L_{\Ga\Ga}$ action it was shown in the work \cite{linden-bell} that under some additional assumptions this integral identity leads to the following integral constraints:
\begin{equation}
    \int_\Sigma d\Sigma_\mu \de T^\mu_\nu V_{(a)}^\nu = \int_{\dd\Sigma} B^{\mu\nu}d\Sigma_{\mu\nu},
\end{equation}
where $V_{(a)}^\mu$ are certain vectors, and $B^{\mu\nu}$ depends on the metric perturbations, $V_{(a)}^\mu$ and their first derivatives. It is important, that $\de T^\mu_\nu$ originates entirely from $X^\rho$. These constraints allow one to reduce Sachs-Wolfe effect on angular CMB fluctuations, as it was shown in \cite{Traschen, TraschenErdley} (see also short review \cite{DeruelleUzan}).\par

The same logic can be applied for theories with change of variables considered above, but the specific calculations will be published in another paper. But even at first glance it follows from \eqref{XAfterChange}, that these integral constraints may gain new contributions to the $X^\mu$ and therefore they will influence resulting Sachs-Wolfe effect. Here we just want to comment on the simplest case of mimetic gravity. Namely, we argue, that in this case $I^\rho$ depends on the perturbed metric and $\xi^\mu$ similarly to GR case. Indeed, from the previous paragraph it follows that for mimetic gravity $X^\mu$ doesn't change under change of variables, and thus it will give the same contribution $\de T^\rho_\mu\xi^\mu$ to the current \eqref{RelativeBackgroundCurrent}. On the other hand, if one use the equations of motion for perturbed change fields $\pp_A$, it is clear from \eqref{LambdaTheoryCurrentCorrectedFinalSV} and the discussion from the start of subsection \ref{CorrectedSuperpotsSubsection}, that superpotential $V_K^{\al\rho}$ for $J^\rho$ will be the same as in GR with Lagrangian \eqref{grLagrBackground}. For $L_{\Ga\Ga}$ it equals to Katz superpotential $V^{\al\rho}_k$, as it was mentioned above. Finally, we note that for mimetic gravity $S^\rho = -X^\rho$, which follows directly from \eqref{SDef}, and therefore $V^{\al\rho}_k-\bar{V}^{\al\rho}_k$ is the superpotential for $I^\rho$. This is precisely what one gets for the usual GR (see eq. (2.37) in \cite{linden-bell}).

\subsection{Link to the Lagrange multiplier formulations}\label{LagrangeMultiplierApproachSection}
Aside from the usual formulation of theories with DFTA, one often considers the alternative formulations with the use of the Lagrange multipliers \cite{mukhanov2014, statja48, statja51, Golovnev201439}. For instance, the simplest variant of such theory for model with DFTA \eqref{lambdaChange} and the Lagrangian \eqref{ScalarTensorLagrangian} is described with the following action:
\begin{equation}
    S[\la_B,\pp_A,\tau^B] = S_{(\la)}[\la_B]+\int d^Dx\;\tau^B\left(\la_B-f_B[\pp_A,\dd_\mu\pp_A]\right),
    \label{LambdaTheoryWithLagrMultiplier}
\end{equation}
where $S_{(\la)}$ is just the action with the Lagrangian \eqref{ScalarTensorLagrangian}, and $\tau^B$ is the Lagrange multiplier. The equations of motion for this theory will be equivalent to those of the considered theory with DFTA \eqref{lambdaChange}. It is not so obvious, though, that the conserved currents are the same. By using the formula \eqref{AntisymmetricSuperpotential} again, one may find that the superpotential for the theory \eqref{LambdaTheoryWithLagrMultiplier} interestingly enough is equal to the corrected superpotential \eqref{LambdaTheoryCurrentCorrectedFinalSV}. Thus, theories \eqref{LambdaLagrangian} and \eqref{LambdaLagrangian} have different expressions for the currents, but they can be made equal by fixing the ambiguity in the form \eqref{LambdaTheoryTildeZChoice}. {This fact is an explanation why the current \eqref{LambdaTheoryCurrentCorrectedFinalSV} does not depend on $N_o$. Furthermore, one may expect result \eqref{LambdaTheoryCurrentCorrectedFinalSV} to hold for more complex cases $N_o > 2$ because it follows from the Lagrange multiplier reformulation.}\par
It is worth noting that the formula \eqref{LambdaTheoryCurrentCorrectedFinalSV} can be obtained from the procedure discussed in \cite{statja46} (for the general case see \cite{ilinPastonUniverse}). In these papers, one start Noether's procedure for the theories with DFTA not from the relation \eqref{NoetherTheoremStep1} but from its original counterpart:
 \begin{equation}
     -\dd_\mu K^\mu = \sum_{i = 0}^{N_{(g)}}Z_{(g)}^{A\al_1\dots\al_i}\dd_{\al_1\dots\al_i}\de g_{\mu\nu},
 \end{equation}
 where $N_{(g)}$ is the order of derivatives in the original metric theory, and $g_{\mu\nu}$ denotes independent metric.
 
\section{Conclusion}\label{Conclusion}
The main contributions of the work are the formulae \eqref{LambdaTheoryCurrentCorrectedFinalSV} for the correct{ed} currents in theories with {DFTA, and also its applications listed in the section \ref{Applications}. Notably, we obtained corrected superpotentials for a general class of Lagrangians that includes mimetic gravity and Regge-Teitelboim theory \eqref{MimeticSuperpotential}, that initially were ill-defined (see discussion in subsection \ref{MimeticSuperpotentialSubsection}). We also established a link between the currents in formulations with DFTA and the currents for their counterparts with Lagrangian multipliers \eqref{LambdaTheoryWithLagrMultiplier}}. Finally, we applied the obtained results to direct physical applications: first law of black holes thermodynamics in theories with change and Traschen-like integral constraints on Sachs-Wolfe effect. In particular, it was shown that for mimetic-like scalar-tensor changes of variables Iyer-Wald entropy can be computed by using the same Komar 2-form as in usual GR. Other cases were also discussed.\par

There are some final remarks, that should be made about these results. First, it is not absolutely clear, do the choices \eqref{LambdaTheoryTildeZChoice},\eqref{ScalarTensorTildeZ3SigmaChoice} and \eqref{ScalarTensorTildeZ2SigmaChoice} uniquely lead to the "good" current $\tilde{J}^\rho$ in the context of discussion after the formula \eqref{JNewIsEqualToJOriginal}. As it is stated in \cite{PetrovCovariantized}, for some theories there exist multiple ways to fix the ambiguity and in the end get the vector density transformation law for $\tilde{J}^\rho$. In our case we additionally demand from the current to obey \eqref{JNewIsEqualToJOriginal}, which decrease the number of the suitable expressions for the current considerably. {One way to possibly reduce it even more is connected with covariant phase space approach (see, for example, \cite{LeeWaldCovariantPS, IyerWaldCovariantPS}). The common point between those results and the conserved currents $J^\rho$ is symplectic potential $\alpha^\mu$. It  (or, more accurately, its restriction to $\de\pp_A$ that originates from the gauge transformations) is defined in \eqref{GeneralActionVar} and is actually a vital part of the Noether's current. In various works \cite{JuliaCovariantPS, HarlowCovariantPS, MargalefVillasenorCovariantPS} it was shown that the boundary conditions at the spatial infinity play crucial role in defining pre-symplectic and symplectic forms on the solution space unambiguously. These methods can be helpful for explaining why (or why not) superpotential \eqref{CorrectedMimeticSuperpotential} is the right choice for Regge-Teitelboim embedding gravity. It is important direction for the future work.}\par
One of the main advantages of the procedure \eqref{NoetherTheoremStep1}-\eqref{JDefinition} is the unique simple formula for conserved current in practically any field theory with continuous symmetry. The ambiguity \eqref{AmbiguityTransform} slightly devalues the original formulae \eqref{JDefinition}, \eqref{AntisymmetricSuperpotential}. However, it is possible, that there exists such modification of this procedure for theories with the higher derivatives, which in the end gives correct results for the currents. As an argument here, one may note, that in some sense the relations \eqref{LambdaTheoryTildeZChoice},\eqref{ScalarTensorTildeZ3SigmaChoice} and \eqref{ScalarTensorTildeZ2SigmaChoice} are obtained by "omitting" the symmetrizers in the standard expressions $Z^{A\rho\al_1\dots\al_k}$. In a very similar way, fixation of the ambiguity is used in the work \cite{PetrovCovariantized} to covariantize the currents. The common thread between \eqref{AmbiguityTransform} and the work \cite{PetrovCovariantized} is the use of non necessarily symmetric $\tilde{Z}^{A\rho\al_1\dots\al_k}$. Possibly enough, this prescription is "correct" in the above sense not only for considered cases but for any field theory. This hypothesis needs deeper analysis and is an important subject of further {research}.\par
Finally, we want to propose another cosmological implication that was not mentioned in the section \ref{CosmologyAppSection}. Namely, there are several powerful approaches that allows one to select cosmological models by analyzing results of Noether's theorem, such as methods developed in works \cite{CapozielloDialektopoulosNoether, CapozzielloLambiaseNoether, CapozzielloNoetherReconstruction}. The key difference from the presented approach is that these methods are usually applied to point-like Lagrangians, usually with the FLRW symmetry imposed on metric in action. Similar results can still be obtained by the formula \eqref{JDefinition}. Moreover, one can also use the formula \eqref{JAmbiguityFixedDefinition} to calculate the corresponding Noether charges. It thus very appealing to construct FLRW-solutions in theories with change by knowing only the solutions from the original theory. In general field theory such program usually cannot be realised. For point-like Lagrangians, however, conservation laws allows one to write first integrals of the corresponing equations of motion. It is very interesting where the application of such methods can lead in the context of the formulae \eqref{LambdaTheoryCurrentCorrectedFinalSV}. Therefore, applications of Noether symmetry approach in the light of the present results are importatnt direction of the future development.

\section*{Acknowledgements}\label{Aknowledgements}
The work of the author is supported by RFBR grant № 20-01-00081. The author is grateful to A.N. Petrov, G. Barnich, and S.A. Paston for the important clarifications and useful discussions.

\section*{Data Availability Statement}\label{DAS}
No Data associated in the manuscript.

\appendix
\section{Appendix}
\subsection{Explicit form of $L_{(k)a}{}^{\rho\al_1..\al_k}$}
The coefficients $L_{(k)a}{}^{\rho\al_1..\al_k}$ defined in the decomposition \eqref{JLDecomposition} plays key role in the calculations throughout the paper. These coefficients are calculated, for example, in \cite{ilinPastonUniverse}: 
\begin{align}
    L_{(k)}{}_a{}^{\rho\ga_1..\ga_k} ={}& \theta\left(N_d-k+1/2\right)\sum_{l = \max(k-N+1,0)}^{\min(k,M)}\Omega{}^{\rho}{}_a{}^{\ga_1..\ga_k}(k-l,l)+ U_a{}^{\rho\ga_1...\ga_k},
    \label{LGeneralExpr}
\end{align}
where
\begin{align}
\nonumber
    \Omega{}^{\rho}{}_a{}^{\al_1..\al_m\be_1..\be_l}(m,l) \equiv& \frac{1}{(m+l)!}\Bigg\{\sum_{j=m+1}^N\sum_{i = 0}^{j-m-1}(-1)^iC^m_{j-i-1}(\dd_{\al_{j-i}..\al_{j-1}}Z^{a\rho\al_1..\al_{j-1}})\times\\&\hspace{4.5cm}\times\dd_{\al_{m+1}..\al_{j-i-1}}H_{(l)}{}_{a\mu}{}^{\be_1..\be_l}\Bigg\}^{\al_1..\al_m\be_1..\be_l},
    \label{OmegaGeneralExpr}
\end{align}
and $U_\al{}^{\rho\ga_1...\ga_k}$ are defined as follows:
\begin{equation}
    K^\rho\equiv -\sum_{i = 0}^CU_a{}^{\rho\al_1\dots\al_i}\dd_{\al_1\dots\al_i}\xi^a.
\end{equation}
Here we used $K^\rho$ defined in \eqref{ActionVariationGeneralCase}.
\subsection{{Commutators for $N_o = 2$ case}}\label{ScalarTensorCommutationRelationsAppendix}
{It is clear that the commutation relations \eqref{LambdaCommutator1}-\eqref{LambdaCommutator3} are still valid for the case $N_o = 2$ considered in \ref{ScalarTensorChangeSubsection}.
Aside from these, there is also need for the commutators of $\dd_{\al\be}$ with $\dd/\dd\dd_\rho\pp_A, \dd/\dd\dd_{\rho\be}\pp_A$ and $\dd/\dd\dd_{\rho\be\ga}\pp_A$. All the needed commutation relations are written down below:
\begin{align}
    &\left[\frac{\dd}{\dd\pp_A},\dd_{\al\be}\right]\la_B = 0,\\
    &\left[\frac{\dd}{\dd\dd_\rho\pp_A},\dd_{\al\be}\right]\la_B = \simdown{\de_\al^\rho\dd_\be\left(\frac{\dd \la_B}{\dd\pp_A}\right)}{\al\be},\\
    &\left[\frac{\dd}{\dd\dd_{\rho\la}\pp_A},\dd_{\al\be}\right]\la_B = \frac{1}{2}\simup{\simdown{\de^\la_\be\dd_\al\left(\frac{\dd \la_B}{\dd\dd_\rho\pp_A}\right)}{\al\be}}{\la\rho}+\frac{1}{2}\frac{\dd \la_B}{\dd\pp_A}\simup{\de^\rho_\al\de^\la_\be}{\rho\la},\label{SecondDerWithSecondJetDerTensorCommutator}\\
    &\left[\frac{\dd}{\dd\dd_{\rho\pp\tau}\pp_A},\dd_{\al\be}\right]\la_B = \frac{1}{6}\simup{\frac{\dd \la_B}{\dd\dd_\rho\pp_A}\de^\pp_\al\de^\tau_\be}{\rho\pp\tau}.
    \label{SecondDerWithThirdJetDerScalarCommutator}
\end{align}}
\subsection{Derivation of the formula \eqref{J'AsDIvergenceWithNon-AntisymmetricSuperpotential}}\label{DerivationOfJ'FormulaAppendix}
To derive \eqref{J'AsDIvergenceWithNon-AntisymmetricSuperpotential} one can use \eqref{KRepresentation} and substitute it to the \eqref{J'Definition}. However, one should first make change $\al_{p+1}\rightarrow\rho$. As the index $\al_{p+1}$ is present in antisymmetrizer, one needs to be careful and should properly split sums before applying the change:
\begin{align}
    &K_{(p)a}{}^{\al_{p+1}\al_1..\al_{p}} = 
    \frac{1}{p+1}\sum_{i = 1}^p\antisimup{K_{(p)a}{}^{\al_{p+1}\al_1..\al_p}}{\al_{p+1}\al_i}-\\
    \nonumber-
    &\sum_{k = p+1}^{N_d}\frac{(-1)^{k-p}}{k+1}\dd_{\al_{p+2}..\al_{k+1}}\antisimup{K_{(k)a}{}^{\al_{p+1}\al_1..\al_p\al_{p+2}..\al_{k+1}}}{\al_{p+1}\al_{k+1}}+\\
    \nonumber
    &+\sum_{k=p+1}^{N_d}\sum_{i = 1}^p\frac{(-1)^{k-p}}{k+1}\dd_{\al_{p+2}..\al_{k+1}}\antisimup{K_{(k)a}{}^{\al_{k+1}\al_1..\al_p\al_{p+1}\al_{p+2}..\al_{k}}}{\al_{k+1}\al_i}+\\
    &+\sum_{k=p+2}^{N_d}\sum_{i = p+2}^k\frac{(-1)^{k-p}}{k+1}\dd_{\al_{p+2}..\al_{k+1}}\antisimup{K_{(k)a}{}^{\al_{k+1}\al_1..\al_{k}}}{\al_{k+1}\al_i}.
\end{align}
The last term equals zero due to the symmetry of the operator $\dd_{\al_{p+2}..\al_{k+1}}$ and antisymmetrization of $K_{(k)a}{}^{\al_{k+1}\al_1..\al_{k}}$ with respect to $\al_{k+1}, \al_i$ for $i\in[p+2,k+1]$. Hence, one can proceed to the change $\al_{p+1}\rightarrow\rho$:
\begin{align}
\nonumber
    &K_{(p)a}{}^{\rho\al_1..\al_{p}} = 
    \frac{1}{p+1}\sum_{i = 1}^p\antisimup{K_{(p)a}{}^{\rho\al_1..\al_p}}{\rho\al_i}-\\
    \nonumber-
    &\sum_{k = p+1}^{N_d}\frac{(-1)^{k-p}}{k+1}\dd_{\al_{p+2}..\al_{k+1}}\antisimup{K_{(k)a}{}^{\rho\al_1..\al_p\al_{p+2}..\al_{k+1}}}{\rho\al_{k+1}}+\\
    \nonumber
    &+\sum_{k=p+1}^{N_d}\sum_{i = 1}^p\frac{(-1)^{k-p}}{k+1}\dd_{\al_{p+2}..\al_{k+1}}\antisimup{K_{(k)a}{}^{\al_{k+1}\rho\al_1..\al_p\al_{p+2}..\al_{k}}}{\al_{k+1}\al_i}.
\end{align}
To simplify this further it is convenient to shift indices for the operators $\dd_{\al_{p+2}..\al_{k+1}}$ (as they are dummy) by $-1$. For convenience we also want to increase inner sum's upper limit to $k-1$ as it does not change anything due to the symmetry of the operator $\dd_{\al_{p+2}..\al_{k+1}}$ ($\dd_{\al_{p+1}..\al_{k}}$ after the mentioned index shift):
\begin{align}
    K_{(p)a}{}^{\rho\al_1..\al_{p}} = &
    \frac{1}{p+1}\sum_{i = 1}^p\antisimup{K_{(p)a}{}^{\rho\al_1..\al_p}}{\rho\al_i}-
    \sum_{k = p+1}^{N_d}\frac{(-1)^{k-p}}{k+1}\dd_{\al_{p+1}..\al_{k}}\antisimup{K_{(k)a}{}^{\rho\al_1..\al_{k}}}{\rho\al_{k}}+\\
    &+\sum_{k=p+1}^{N_d}\sum_{i = 1}^{k-1}\frac{(-1)^{k-p}}{k+1}\dd_{\al_{p+1}..\al_{k}}\antisimup{K_{(k)a}{}^{\al_{k}\rho\al_1..\al_{k-1}}}{\al_{k}\al_i}.
    \label{KGeneralFormWithRho}
\end{align}
Now one can multiply this expression by $\dd_{\al_1..\al_p}\xi^{a}$ and substitute the result into \eqref{J'AsDIvergenceWithNon-AntisymmetricSuperpotential}:
\begin{align}
    -J'^{\rho} =& \sum_{p = 1}^{N_d}\sum_{i = 1}^p\frac{1}{p+1}\antisimup{K_{(p)a}{}^{\rho\al_1..\al_p}}{\rho\al_i}\dd_{\al_1..\al_p}\xi^{a}-\sum_{p = 0}^{N_d}\sum_{k = p+1}^{N_d}\frac{(-1)^{k-p}}{k+1}\dd_{\al_{p+1}..\al_{k}}\antisimup{K_{(k)a}{}^{\rho\al_1..\al_{k}}}{\rho\al_{k}}\times\\
    &\times\dd_{\al_1..\al_p}\xi^{a}+\sum_{p = 0}^{N_d}\sum_{k=p+1}^{N_d}\sum_{i = 1}^{k-1}\frac{(-1)^{k-p}}{k+1}\dd_{\al_{p+1}..\al_{k}}\antisimup{K_{(k)a}{}^{\al_{k}\rho\al_1..\al_{k-1}}}{\al_{k}\al_i}\dd_{\al_1..\al_p}\xi^{a}.
    \label{J'ThroughKIdentities}
\end{align}
Our next step is to convert r.h.s to the full derivative. We start with the first two terms by changing the summation order in the second term:
\begin{equation}
    \sum_{p = 0}^{N_d-1}\sum_{k = p+1}^{N_d} = \sum_{k = 1}^{N_d}\sum_{p = 0}^{k - 1}.
    \label{pkSummChange}
\end{equation}
After that one should interchange indices $k\leftrightarrow p$ in the second term and change them$i\rightarrow k$ in the first one. Finally, after applying shift $k\rightarrow k+1$ in the second term the can be properly merged:
\begin{align}
\nonumber
    -J'^{\rho} =& 
    \sum_{p = 1}^{N_d}\sum_{k = 1}^p\frac{1}{p+1}\left[\antisimup{K_{(p)a}{}^{\rho\al_1..\al_p}}{\rho\al_k}\dd_{\al_1..\al_p}\xi^{a}+(-1)^{p-k}\dd_{\al_k..\al_{p}}\antisimup{K_{(p)a}{}^{\rho\al_1..\al_p}}{\rho\al_p}\dd_{\al_1..\al_{k-1}}\xi^{a}\right]+\\
    &+\sum_{p = 0}^{N_d}\sum_{k=p+1}^{N_d}\sum_{i = 1}^{k-1}\frac{(-1)^{k-p}}{k+1}\dd_{\al_{p+1}..\al_{k}}\antisimup{K_{(k)a}{}^{\al_{k}\rho\al_1..\al_{k-1}}}{\al_{k}\al_i}\dd_{\al_1..\al_p}\xi^{a}.
    \label{J'IdentitiesGroupFirstTerms}
\end{align}
First term can be further simplified by subsequently using Leibniz' rule in the form similar to \eqref{Leibniz'Rule}:
\begin{equation}
\begin{split}
&(-1)^{p-k}\dd_{\al_k..\al_{p}}(Y^{\al_1..\al_p})\dd_{\al_1..\al_{k-1}}Q=\\
&\hspace{4cm}=\sum_{i = k}^p(-1)^{p-i}\dd_{\al_i}\left(\dd_{\al_{i+1}..\al_p}Y^{\al_1..\al_p}\dd_{\al_1..\al_{i-1}}\xi^{\mu}\right) - Y^{\al_1\dots\al_p}\dd_{\al_1..\al_p}Q,
    \label{DerivativePartialCaseRelation}
\end{split}
\end{equation}
which is true for the arbitrary functions $Y^{\al_1..\al_p}$ (even without any index symmetry) and the arbitrary function $Q$. Then by using the simple relation:
\begin{equation}
    \antisimup{K_{(p)a}{}^{\rho\al_1..\al_p}}{\rho\al_p}\dd_{\al_1..\al_{p}}\xi^{a} = \antisimup{K_{(p)a}{}^{\rho\al_1..\al_p}}{\rho\al_k}\dd_{\al_1..\al_{p}}\xi^{a},\;\; k\leq p,\;\; p\geq 1,
\end{equation}
one may rewrite \eqref{J'IdentitiesGroupFirstTerms} in the following form:
\begin{align}
\nonumber
    -J'^{\rho} =& 
    \sum_{p = 1}^{N_d}\sum_{k = 1}^p\sum_{i= k}^p\frac{(-1)^{p-i}}{p+1}\dd_{\al_i}\left(\dd_{\al_{i+1}..\al_p}\antisimup{K_{(p)a}{}^{\rho\al_1..\al_p}}{\rho\al_p}\dd_{\al_1..\al_{i-1}}\xi^{a}\right)+\\
    &+\sum_{p = 0}^{N_d}\sum_{k=p+1}^{N_d}\sum_{i = 1}^{k-1}\frac{(-1)^{k-p}}{k+1}\dd_{\al_{p+1}..\al_{k}}\antisimup{K_{(k)a}{}^{\al_{k}\rho\al_1..\al_{k-1}}}{\al_{k}\al_i}\dd_{\al_1..\al_p}\xi^{a}.
    \label{J'IdentitiesFirstTermsDerivative}
\end{align}
The same trick with Leibniz' rule can be applied to the second term of \eqref{J'IdentitiesFirstTermsDerivative}, which will lead to the terms that are full derivatives and the following contribution:
\begin{equation}
    \sum_{p = 0}^{N_d}\sum_{k=p+1}^{N_d}\sum_{i = 1}^{k-1}\frac{(-1)^{k-p}}{k+1}\antisimup{K_{(k)a}{}^{\al_{k}\rho\al_1..\al_{k-1}}}{\al_{k}\al_i}\dd_{\al_1..\al_{k}}\xi^{a}.
\end{equation}
Note, that the minimal number of derivatives in $\dd_{\al_1..\al_k}\xi^a$ is greater or equal than $2$.
It becomes obvious then that this expression is zero if one takes into account antisymmetrizer and the symmetry of $\dd_{\al_1..\al_{k}}\xi^{a}$. Thus, the formula \eqref{J'IdentitiesFirstTermsDerivative} can be written as follows:
\begin{align}
\nonumber
    &-J'^{\rho} = 
    \sum_{p = 1}^{N_d}\sum_{k = 1}^p\sum_{i= k}^p\frac{(-1)^{p-i}}{p+1}\dd_{\al_i}\left(\dd_{\al_{i+1}..\al_p}\antisimup{K_{(p)a}{}^{\rho\al_1..\al_p}}{\rho\al_p}\dd_{\al_1..\al_{i-1}}\xi^{a}\right)-\\
    &-\sum_{p = 0}^{N_d}\sum_{k=p+1}^{N_d}\sum_{i = 1}^{k-1}\sum_{s = p+1}^k\frac{(-1)^{k+s}}{k+1}\dd_{\al_s}\left(\dd_{\al_{s+1}..\al_k}\antisimup{K_{(k)a}{}^{\al_{k}\rho\al_1..\al_{k-1}}}{\al_{k}\al_i}\dd_{\al_1..\al_{s-1}}\xi^{a}\right).
    \label{J'IdentitiesFullDivergence}
\end{align}
As terms in the first and the second sequences of sums do not depend on $k$ and $p$ respectively, one can change summation order to calculate sums over these indices. This can be easily done for the first sequence of sums by using the following relation:
\begin{equation}
\sum_{k = 1}^p\sum_{i = k}^p = \sum_{i = 1}^p\sum_{k = 1}^i
\label{skIndexChange}
\end{equation}
For the second term in \eqref{J'IdentitiesFullDivergence} one should set the upper bound of sum over $p$ to $N_d-1$ (because the sum over $k$ will give zero for $p = N_d$) and use \eqref{pkSummChange}, then perform index shift $p\rightarrow p+1$ and use \eqref{skIndexChange}. The result will be the following:
\begin{align}
\nonumber
    -J'^{\rho} = 
    &\sum_{p = 1}^{N_d}\sum_{k= 1}^p\frac{k(-1)^{p-k}}{p+1}\dd_{\al_k}\left(\dd_{\al_{k+1}..\al_p}\antisimup{K_{(p)a}{}^{\rho\al_1..\al_p}}{\rho\al_p}\dd_{\al_1..\al_{k-1}}\xi^{a}\right)+\\
    &-\sum_{k=1}^{N_d}\sum_{i = 1}^{k-1}\sum_{s = 1}^k\frac{s(-1)^{k-s}}{k+1}\dd_{\al_{s}}\left(\dd_{\al_{s+1}..\al_k}\antisimup{K_{(k)a}{}^{\al_{k}\rho\al_1..\al_{k-1}}}{\al_{k}\al_i}\dd_{\al_1..\al_{s-1}}\xi^{a}\right).
    \label{J'IdentitiesSummSimplified}
\end{align}
One can now put the derviatives $\dd_{\al_k}$ and $\dd_{\al_s}$ in the first and the second terms respectively out of the sums by renaming these indices in each term of the sums to $\be$. As it was for \eqref{KGeneralFormWithRho}, one should be careful when doing this because the indices $\al_k$ and $\al_s$ can be encountered as the indices of the antisymmetrizer. The final answer is given by the formula:
\begin{align}
\nonumber
    -J'^{\rho} = 
    &\dd_{\be}\Bigg[\sum_{p = 2}^{N_d}\sum_{k= 1}^{p-1}\frac{k(-1)^{p-k}}{p+1}\dd_{\al_k..\al_{p-1}}\left(\antisimup{K_{(p)a}{}^{\rho\be\al_1..\al_{p-1}}}{\rho\al_{p-1}}+\antisimup{K_{(p)a}{}^{\be\rho\al_1..\al_{p-1}}}{\be\al_{p-1}}\right)\times\\
\nonumber
    &\times\dd_{\al_1..\al_{k-1}}\xi^{a}+\sum_{p = 1}^{N_d}\frac{p}{p+1}\antisimup{K_{(p)a}{}^{\rho\be\al_1..\al_{p-1}}}{\rho\be}\dd_{\al_1..\al_{p-1}}\xi^{a}
    -\\
\nonumber
    &-\sum_{p=3}^{N_d}\sum_{i = 1}^{p-2}\sum_{s = 1}^{p-1}\frac{s(-1)^{p-s}}{p+1}\dd_{\al_{s}..\al_{p-1}}\antisimup{K_{(p)a}{}^{\al_{p-1}\rho\be\al_1..\al_{p-2}}}{\al_{p-1}\al_i}\dd_{\al_1..\al_{s-1}}\xi^{a}-\\
    &\hspace{5cm}-\sum_{p=2}^{N_d}\frac{p(p-1)}{p+1}\antisimup{K_{(p)a}{}^{\be\rho\al_1..\al_{p-1}}}{\be\al_{p-1}}\dd_{\al_1..\al_{p-1}}\xi^{a}\Bigg].
    \label{J'IdentitiesSummSimplifiedFinal}
\end{align}
One can easily compare this result with \eqref{J'AsDIvergenceWithNon-AntisymmetricSuperpotential} and \eqref{NonSymmetricAddition} to ensure that they coincide.
\end{document}